\documentclass[apj]{emulateapj}

\usepackage{graphicx}
\usepackage{times}
\usepackage{dcolumn}
\usepackage{natbib}
\usepackage{apjfonts}

\newcommand{\mbh}{\ensuremath{M_{\mathrm{BH}}\,}}

\newcommand{\ratio}{\ensuremath{M_{\mathrm{BH}} - M_{*,\mathrm{Bulge}}\,}}
\newcommand{\mbhtot}{\ensuremath{M_{\mathrm{BH}} - M_{*,\mathrm{Total}}\,}}

\slugcomment{Accepted for publication in The Astrophysical Journal on December 6, 2012}

\shorttitle{}
\shortauthors{M. Schramm et al.}
\begin{document}
\title{The black hole - bulge mass relation of Active Galactic Nuclei 
in the Extended $Chandra$ Deep Field - South Survey}
\author{Malte Schramm\altaffilmark{1} and John D. Silverman}
\affil{Kavli Institute for the Physics and Mathematics of the Universe,  Todai Institutes for Advanced Study, the University of Tokyo,  Kashiwa, Japan 277-8583 (Kavli IPMU, WPI)}

\altaffiltext{1}{malte.schramm@ipmu.jp}

\begin{abstract}

We present results from a study to determine whether relations, established in the local Universe, between the mass of supermassive  
black holes (SMBHs) and their host galaxies are in place at higher  
redshifts.  We identify a well-constructed sample of 18 X-ray-selected, broad-line Active Galactic Nuclei (AGN) in the  
Extended $Chandra$ Deep Field South - Survey with $0.5 < z < 1.2$. This redshift range is chosen to ensure that HST imaging is available with  
at least two filters that bracket the 4000~\AA\, break thus providing  
reliable stellar mass estimates of the host galaxy by accounting for  
both young and old stellar populations. We compute single-epoch, virial black hole masses from  
optical spectra using the broad MgII emission line. For essentially all galaxies in our sample, 
their total stellar mass content agrees remarkably  
well, given their BH masses, with local relations of inactive galaxies and active SMBHs. 
We further decompose the total stellar mass into bulge and disk components separately with full knowledge of the HST point-spread-function.
We find that $\sim80$\% of the sample is consistent with the local \ratio relation even with $72\%$ 
of the host galaxies showing the presence of a disk.
In particular, bulge dominated hosts are more aligned with the local relation than those with
prominent disks. We further discuss the possible physical mechanisms that are capable  
building up the stellar mass of the bulge from an extended disk of  
stars over the subsequent eight Gyrs.

\end{abstract}


\section{Introduction}\label{sec:intro}

A determination of the physical mechanisms through which supermassive black holes are built up at the centers of
galaxies have been one of the key issues in astrophysics \citep[see][]{Kormendy(1995)}.
Such processes are thought to further provide a link black hole growth and the formation
of the bulges of their host galaxies based on both observations and theory.  Correlations between the mass of the central black hole and absolute magnitude \citep{Magorrian(1998),Marconi(2003),Haering(2004)}, 
and/or stellar velocity dispersion \citep{Gebhardt(2000),Merritt(2001)} of the spheroidal component indicate that the mass ratio between a SMBH and its bulge 
is constant over a wide dynamic range in mass (e.g.  $M_{BH}/M_{Bulge}$= 0.0014;
\cite{Haering(2004)}; hereafter $M_{BH}-M_{Bulge}$ relation). We will refer to this relation as the local relation.

Over the past years, several studies have addressed whether there is an evolution of the mass relations between the 
central black hole and its host galaxy. 
Such studies must rely upon galaxies with accreting SMBHs (i.e.,  
Active Galactic Nuclei; AGN) since the region of influence surrounding  
black holes cannot be resolved at higher redshifts.  While those  
hidden by obscuration (i.e., type 2 AGNs) give a rather clean view of  
their host galaxy, unobscured (type 1) AGN are the only systems for  
which black hole masses can be measured.  Although,  an estimate of  
the mass of the host bulge is challenging due to the glare of a luminous AGN  
that only gets more difficult at high redshift.  Fortunately, optical  
imaging from space with HST can be used to disentangle the light  
between an AGN and its host galaxy \citep[e.g.,][]{Sanchez(2004),Jahnke(2004),Jahnke(2009),Bennert(2011),Cisternas(2011)} due to the high spatial resolution  
and well understood point spread function.  Alternatively, it is also  
possible to measure the stellar velocity dispersion from optical  
spectra for less luminous AGNs \citep{Woo(2008)};  this method  
requires high signal-to-noise spectra that limits its application to  high redshift AGNs.

 Even if the host galaxy is resolved only limited spectral coverage is usually available to estimate 
stellar masses. Single-band studies are therefore restricted to the black hole mass - luminosity relation or have to make assumptions on the mass-to-light ratio of the host galaxy 
\citep[see][]{Peng(2006),Peng(2006)2,Decarli(2010)2,Decarli(2010)}. \cite{Merloni(2010)} implemented a new approach to measure the stellar mass content of AGN host galaxies through template fitting of 
the broad-band photometric spectral energy distribution \citep[][]{Brusa(2009),Xue(2010)}.  With this approach, \citet{Merloni(2010)} estimate the total stellar mass content which provides only an upper 
limit to the bulge mass. Bennert et al. (2011) take a significant step forward by using the  
multi-band HST data available in the GOODS \citep{Giavalisco(2004)} fields to decompose the AGN  
and host galaxy light including a bulge component tractable through multiple filter bandpasses.  Unfortunately, the  
sample is selected to be a redshifts ($1 < z < 2$) for which the  
optical imaging falls below the rest-frame 4000 \AA\, break. Surprisingly, the aforementioned studies find elevated black hole  
masses as compared to either the bulge component \citep{Woo(2008),Bennert(2011)} or total \citep{Merloni(2010)} stellar mass of  
their host galaxy.  Recently, \citet{Jahnke(2009)} and \citet{Cisternas(2011)} report that the mass ratio  
between the black hole and the total stellar mass of its host galaxy  
is similar to local values possibly indication of an undermassive bulge.

Even with the considerable effort achieved to date, there are  
several challenges that need to be met in order to accurately  
determine the evolution of the \ratio at higher redshift. First, the  
decomposition of optical light is more difficult due to the strong  
surface brightness dimming of the host galaxy as compared to the AGN.   
To mitigate this effect, high resolution imaging with high signal-to- noise is needed to adequately resolve the host galaxy especially for  
bright AGN.  Equally important, at least, one rest-frame optical color  
and a luminosity is needed to constrain the  
stellar mass content of the host galaxy \citep{Bell(2003)}.  A color that covers the  
4000 \AA~break provides a good estimator on the underlying stellar mass-to-light ratio.  It is worth highlighting that the 4000 \AA\, break moves  
out of the optical filter bands at $z>1.2$ thus requiring deep high-resolution NIR imaging. Furthermore, due to the limited physical resolution at high  
redshift and the fact that galaxies become more compact,  it may be challenging to classify galaxies morphologically  
such as distinguishing between disturbed and undisturbed hosts. 

A determination of the \ratio~relation using AGN samples, also requires an assessment of the possible biases originating from 
selection of AGN \citep[see][]{Salviander(2007),Lauer(2007)}. While quiescent galaxies are selected by their magnitude or
luminosity, active galaxies (e.g., unobscured, broad-line AGN) are often selected by their optical nuclear luminosity or magnitude. The bias introduced by the luminosity (i.e., mass) limit will have a stronger effect at the high mass end of the black hole mass 
function which is strongly decreasing.
Offsets from the local \ratio relation seen in samples of luminous AGN with massive BHs \citep[e.g.,][]{Merloni(2010),Bennert(2011),Peng(2006),Peng(2006)2} may be explained by such a bias. Therefore, a sample selected at lower luminosities that fall well below the knee of the black hole mass function should be less impacted by such a bias. 

In this study, we determine the \mbhtot  and \ratio   
relations at $0.5<z<1.2$ using a sample of 18 X-ray
selected broad-line AGN (BLAGN) from the Extended $Chandra$ Deep Field - South Survey.  
Based on HST/ACS imaging from GEMS \citep{Rix(2004)} and GOODS \citep{Giavalisco(2004)}, we measure the stellar  
mass content of their host galaxies including the bulge component.  We  
specifically focus on this redshift range so that there is at least  
one HST band above and below the 4000~\AA~break thus providing a rest-frame color required for accurate conversion of light to mass.  Black hole masses are determined using single epoch virial mass estimation  
based on the MgII emission line.  In Section 2, we describe our  
sample. In Section 3, we describe our analysis of the HST/ACS data that involves the image decomposition of the total light into AGN, bulge and disk decomposition, and stellar mass estimation.
Black hole masses are fully detailed in Section 4. Section 5 and 6 presents the results including a discussion of the relations between the mass of the SMBH and their total/bulge stellar mass.
Finally, in Section 7 we give a summary of the results. Throughout this paper we assume a flat cosmology with $H_0=70~\mathrm{ km\,s}^{-1}\,\mathrm{Mpc}^{-1}$, $\Omega_M=0.3$ and
$\Omega_{\Lambda}=0.7$.

\begin{figure}
\begin{center}
 \includegraphics[clip,angle=0,width=245pt]{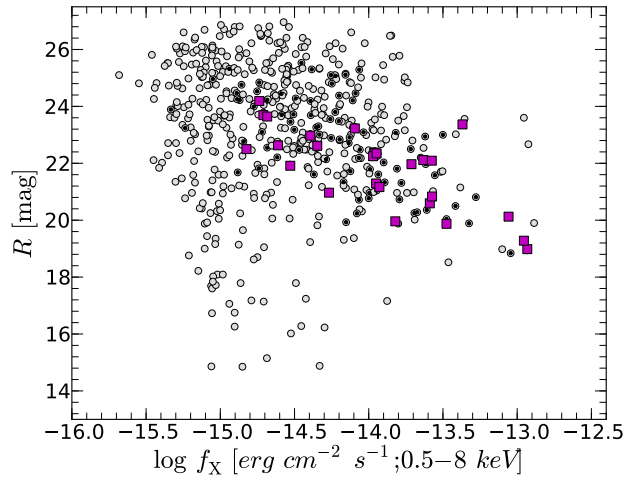}
 \caption{Optical $R$-band magnitude versus broad-band X-ray flux (0.5-8 keV; units of ergs cm$^2$ s$^{-1}$) for all X-ray sources from Lehmer et al. 2005 (grey circles). 
All sources classified as BLAGN have been marked with a black dot. Our final sample selection is shown as magenta solid squares. Objects falling below the typical
relation for type I AGNs are the obscured (type II) AGNs \citep[see Figure 5 of][]{Silverman(2010)}.}
\label{ecdfs_distr}
\end{center}
\end{figure}

\section{AGN Sample}

Currently, broad-line (type 1) AGNs provide the only means to establish the relation between BH mass  
and galaxy mass beyond the local universe.  This is due to  
the fact that BH mass measurements rely upon a determination of the  
velocity widths of gas in the vicinity of the BH as provided by broad  
emission lines \citep[e.g.,][]{Kaspi(2000),Vestergaard(2006)}. High-resolution imaging (best if taken  
from space) can then be used to detect the extended emission from the underlying host  
galaxy. There have been numerous studies of the host galaxies of type 1 AGNs  
using such techniques \citep[e.g.,][]{Jahnke(2004),Jahnke(2009),Sanchez(2004),Bennert(2011),Cisternas(2011)}.
 
We aim to take advantage of type 1 AGNs that are found in X-ray surveys,  
such as the $Chandra$ Deep Field South - Survey, that reach faint  
depths.  These X-ray sources are likely to have a wide range in their  
optical properties that includes those of lower luminosity both missed  
in optically-selected samples such as SDSS, and more favorable for the  
study of their host galaxy.  There are numerous papers on the host  
galaxies of  X-ray selected AGN that may be of interest to the reader  
\citep[e.g.,][]{Grogin(2005),Pierce(2007),Ammons(2009),Silverman(2008)}.
 
Here, we specifically select type 1 AGNs from the compilation of  
\cite{Silverman(2010)} that provide spectroscopic redshifts and  
classification of X-ray sources in the the Extended $Chandra$ Deep Field  
South - Survey \citep{Lehmer(2005)}. These are objects with at least  
one broad emission line having a FWHM greater than 2000 km s$^{-1}$.   
We further require that an available spectrum has a good enough  
quality to perform our emission line fitting procedure to estimate  
virial black hole masses.

 We then demand that each type 1 AGN has been observed by HST.  The  
ECDFS is covered by the GEMS \citep{Rix(2004)} and GOODS \citep{Giavalisco(2004)} surveys  
in the central area. GEMS consists of imaging in two optical HST  
filters (ACS F606W and F850LP) while GOODS has four filters (ACS F435W,  
F606W, F775W and F850LP). Unfortunately, some sources are located on  
the outskirts of the ECDFS and therefore no HST coverage is  
available. Even though, extensive ground based data is available of  
the full ECDFS area from various observing campaigns \citep[e.g., MUSYC survey;][]{Cardamone(2010)}, we choose  
to avoid any biases that may appear due to the inclusion low  
resolution data.  We further stress that is essential to have at  
least two filters that bracket the 4000 \AA\,break in the rest-frame of  
the host galaxy for accurate estimation of the mass-to-light ratio  
and the stellar mass (see the following section).  To do so, we elect  
to restrict the type 1 AGN sample to $0.5 < z < 1.2$ that allows us to  
determine accurate rest-frame B-V colors for the entire sample.  In  
addition, we apply the same selection to the deeper 2Msec catalog (Luo  
et al. 2008) and identify one additional source.  Our final sample  
consists of 18 type 1 AGN with half falling in the GOODS area and the other half within the GEMS field.   In Figure 1, we show the distribution of X-ray flux and R-band optical magnitude of $Chandra$ sources and highlight those within our type 1 AGN sample.  It is apparent that the sample spans about two dex in both X-ray flux or luminosity, and optical brightness. The final sample covers the full region of the $f_X-R$ plane as the overall BLAGN sample.

\begin{table*}
\scriptsize
      \caption{{\bf Sample:} Target List and results from the analysis. The different columns show
(1) Object ID taken from the Lehmer et al., (2),(3) Ra DEC coordinates, (4) redshift, (5) absolute V-bband magnitude,(6) rest-frame U-V color of the host galaxy
,(7) Sersic Index from single Sersic fit,(8) half-light radius of the bulge and disk in arcsec in F850LP (9) total stellar mass, (10) bulge-to-total luminosity ratio, (11) FWHM of the MgII emission line,
(12) continuum luminosity at 3000\AA, (13) BH mass, (14) Eddington Ratio, (15) nuclear-to-host ratio in F850LP, (16) survey field:GEMS [1] \& GOODS [2]}
         \label{T:obs}
\begin{center}
\begin{tabular}{llllllllllllllllllllllllllllllll}
\hline\noalign{\smallskip}
    Source ID   & RA          & Dec      & $z$   & $M_V^{\rm{host}}$ & $U-V$ &  Sersic & $r_\mathrm{bulge}/r_\mathrm{disk}$& $M_{\rm{tot}}$ [$M_{\odot}$] & B/T & FWHM\footnotemark[3] & $L_{3000}$ & log \mbh & $\epsilon$&N/H &Field\\
   $[1]$         & [2]         & [3]      & [4]   & [5]               & [6]   &  Index [7]          & [8]                          & [9] & [10] & [11] & [12]& [13]&[14]&[15]&[16]\\
    
\noalign{\smallskip}\hline\noalign{\smallskip}
158                 &  52.942418 &  -27.952407 &  0.717 &  -21.70\footnotemark[2] &  0.78 &  4.0 & 0.83/- &  10.66 &  1.00 &  3.24 &  43.22 &  7.08 &  0.07 &0.05&[1]\\
170                 &  52.949839 &  -27.845910 &  1.065 &  -20.33\footnotemark[2] &  0.04 &  2.2 & 0.07/0.12 &  9.80 &  0.50 &  3.12 &  42.85 &  6.98 &  0.05&0.11&[1]\\
250                 &  53.001530 &  -27.722073 &  1.037 &  -21.20\footnotemark[2] &  0.84 &  7.7 & 0.13/- &  10.19 &  1.00 &  4.12 &  44.01 &  7.64 &  0.09&1.00&[2]\\
271\footnotemark[1] &  53.125255 &  -27.756535 &  0.960 &  -21.08                 &  0.76 &  1.4 & -/- &  10.49 &  0.23 &  2.72 &  43.86 &  7.26 &  0.19&0.54&[2]\\
273 		    &  53.016951 &  -27.623708 &  0.970 &  -20.39\footnotemark[2] &  0.96 &  4.3 & 0.22/- &  10.29 &  1.00 &  6.96 &  43.84 &  8.13 &  0.03&1.22&[1]\\
305                 &  53.036121 &  -27.792822 &  0.544 &  -22.47\footnotemark[2] &  0.58 &  4.0 & 0.77/- &  11.01 &  1.00 &  5.80 &  44.98 &  8.52 &  0.14&1.51&[2]\\
333                 &  53.057789 &  -27.602162 &  1.044 &  -21.06\footnotemark[2] &  0.41 &  1.1 & 0.08/0.14 &  10.33 &  0.35 &  6.40 &  43.29 &  7.80 &  0.02&0.19&[1]\\
339                 &  53.062421 &  -27.857514 &  0.675 &  -22.00\footnotemark[2] &  1.04 &  5.5 & 0.32/- &  10.96 &  1.00 &  9.81 &  42.60 &  7.74 &  0.00&<0.03&[2]\\
348                 &  53.071454 &  -27.717531 &  0.569 &  -21.51\footnotemark[2] &  0.72 &  1.2 & 0.21/0.24 &  10.55 &  0.40 &  5.82 &  43.91 &  7.78 &  0.04&0.08&[2]\\
375                 &  53.110383 &  -27.676530 &  1.031 &  -22.41                 &  -0.12&  1.1 & -/- &  10.55 &  0.10 &  3.02 &  45.18 &  7.92 &  0.63&3.14&[2]\\
379                 &  53.112521 &  -27.684732 &  0.737 &  -22.49                 &  0.02 &  2.4 & -/- &  10.66 &  0.32 &  10.30 &  45.05 &  9.04 &  0.05&3.28&[2]\\
413                 &  53.156053 &  -27.666706 &  0.664 &  -20.40\footnotemark[2] &  0.81 &  1.9 & 0.30/0.36 &  10.19 &  0.41 &  2.29 &  43.38 &  6.95 &  0.16&0.29&[2]\\
417                 &  53.158807 &  -27.662460 &  0.837 &  -20.36\footnotemark[2] &  1.02 &  2.1 & 0.10/0.37 &  10.39 &  0.33 &  5.22 &  44.67 &  8.25 &  0.12&5.50&[2]\\
465                 &  53.199639 &  -27.696625 &  0.740 &  -21.60 \footnotemark[2] &  0.87 &  2.2 & 0.40/0.51  &  10.76 &  0.24 &  5.60 &  43.78 &  7.92 &  0.04&0.22&[1]\\
516                 &  53.246074 &  -27.727665 &  0.733 &  -22.48	          &  0.46 &  2.3 & -/- &  10.88 &  0.67 &  6.08 &  43.44 &  7.80 &  0.02&0.08&[1]\\
540                 &  53.256378 &  -27.761801 &  0.622 &  -22.03\footnotemark[2] &  0.79 &  1.6 & 0.32/0.67 &  10.88 &  0.49 &  5.18 &  43.07 &  7.52 &  0.02&<0.03&[1]\\
597                 &  53.292483 &  -27.811635 &  1.034 &  -21.93\footnotemark[2] &  0.92 &  2.8 & 0.27/0.66 &  10.91 &  0.77 &  7.22 &  43.54 &  8.01 &  0.02&0.18&[1]\\
712                 &  53.370529 &  -27.944765 &  0.841 &  -22.98\footnotemark[2] &  0.96 &  2.3 & 0.21/0.96 &  11.34 &  0.70 &  6.44 &  44.87 &  8.55 &  0.10&0.99&[1]\\

\noalign{\smallskip}\hline
\footnotetext{[1] ID taken from Giacconi et al. 2002}
\footnotetext{[2] direct bulge estimate through either B or B+D fit to imaging data}
\footnotetext{[3] units of 1000 km/s}
\end{tabular}
\end{center}
\end{table*}

\section{Observed Host Galaxy Properties}

\subsection{AGN-host decomposition and bulge correction}

 The first step to obtain information on the host galaxy is to remove  
 the contribution of the AGN component from the broad-band
 HST images. The separation of galaxy light from that of the  
 nuclear point source in luminous AGN is challenging
 even at lower redshifts since the AGN can outshine the host galaxy by  
 several magnitudes. Objects in our study have lower luminosities (due to their X-ray selection with deep observations) thus the contamination from the point source is substantially weaker compared to similar studies using optically-seleted quasars. 

 Knowledge of the point spread function (PSF) at the position of the
 AGN is crucial for our further analysis. The ACS PSF in the GEMS  
 survey is known to vary across the field \citep{Jahnke(2004)}. Therefore, we create a local PSF for each AGN by averaging all  
 stars within a radius of  60 arcsec around the target. Each high
 S/N PSF consists of about 30-40 stars. The remaining uncertainties  
 between individual stars are included in the variance frame as
 an additional contribution from the rms image of each PSF. In the  
 GOODS fields, the estimation of a proper PSF is more difficult.
 Each tile has only a limited number of unsaturated stars (5- 20) with strong spatial variations, in some cases, between stars located
 in the center and at the edges of each tile. For most of our objects,  
 we used the same strategy as for GEMS by creating a local
 mean PSF for each AGN position. In three cases, the AGN was close to the edge of the tile thus we used the nearest  
 star as our PSF reference.
 
\begin{figure*}
\begin{center}
\includegraphics[clip,angle=0,width=380pt]{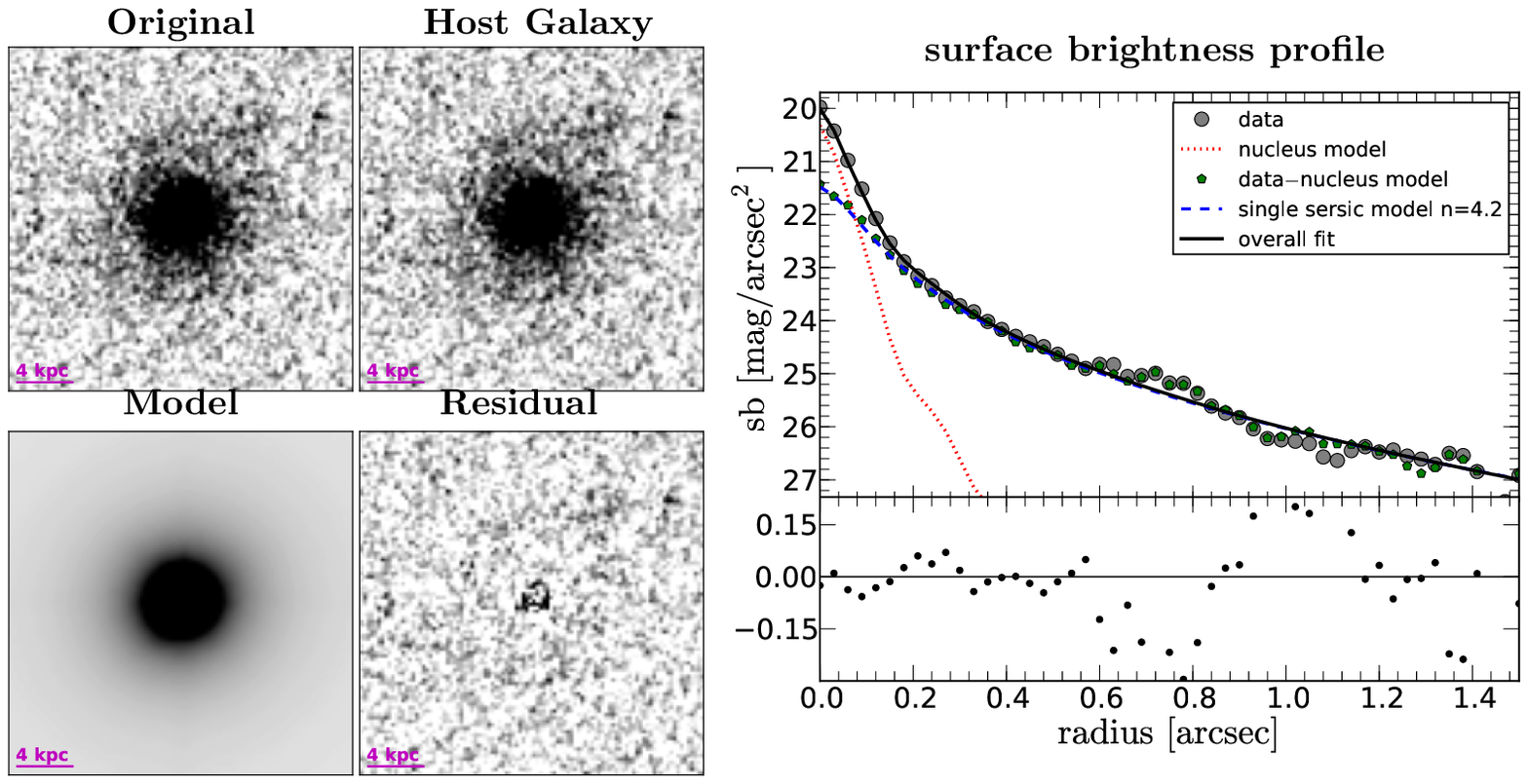}
\includegraphics[clip,angle=0,width=380pt]{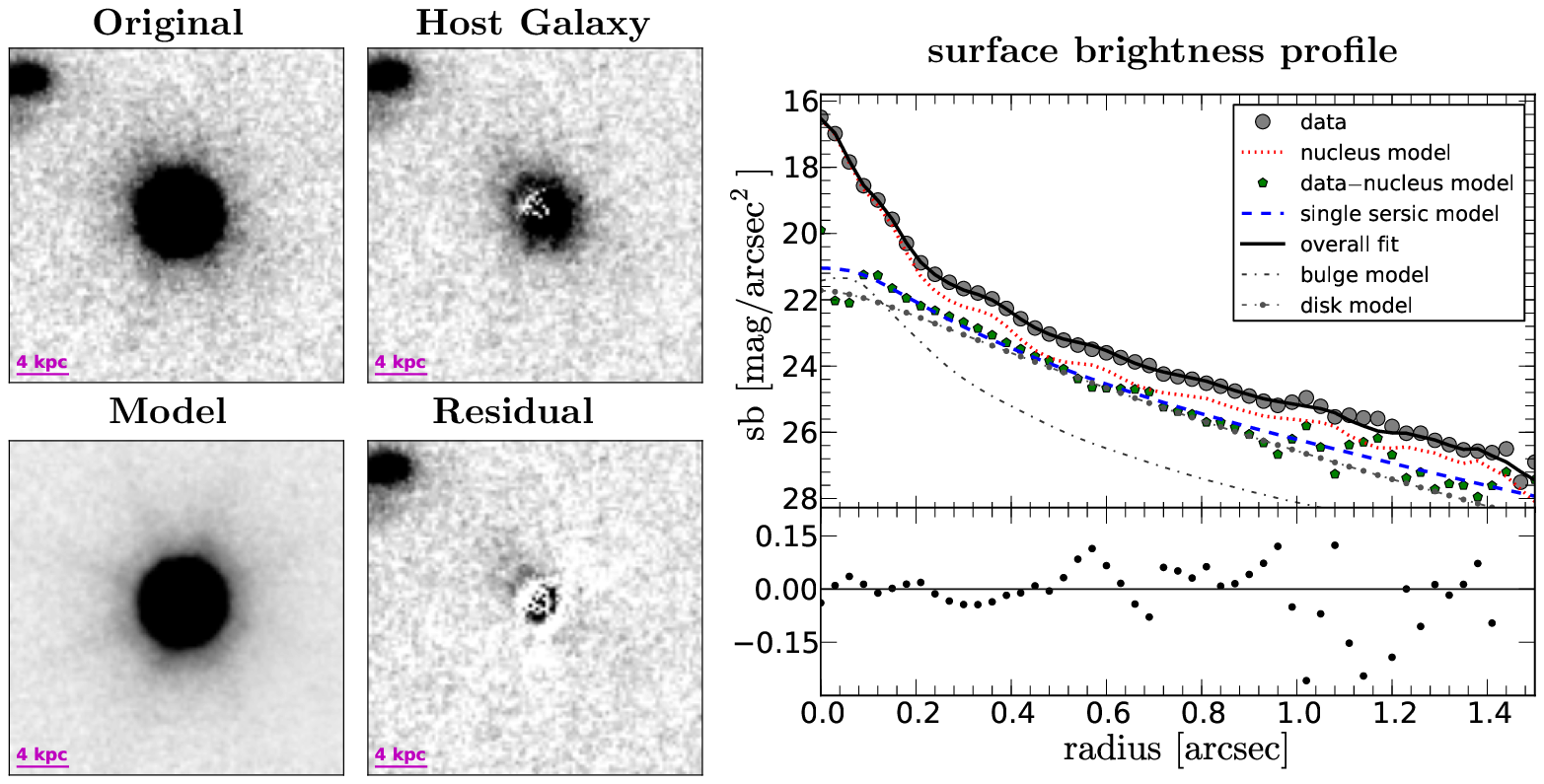}
\includegraphics[clip,angle=0,width=380pt]{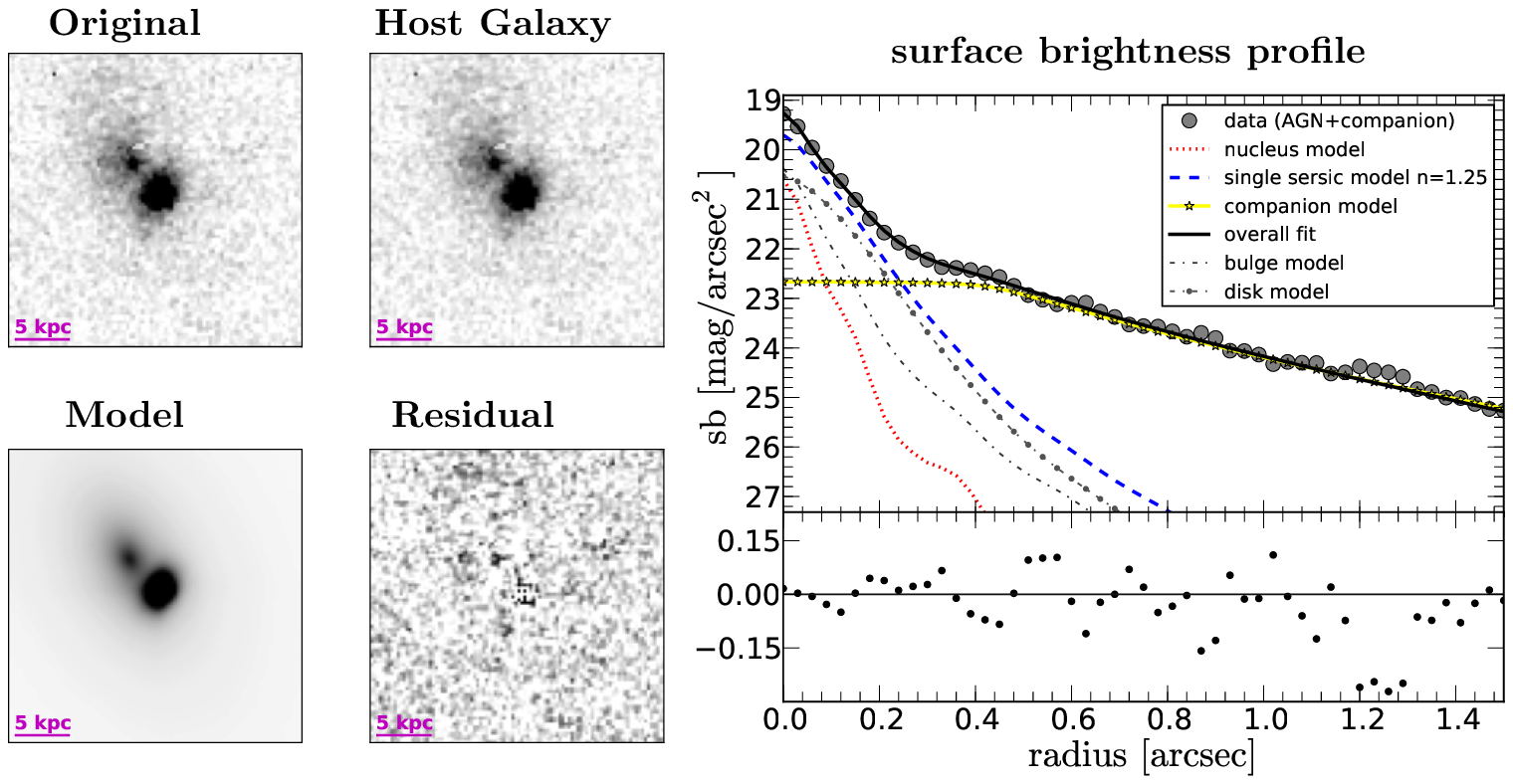}
 \caption{Three examples of HST image decomposition and surface profile fitting.  {\bf Top panels:} AGN ID-158 having a bulge-to-total (B/T) light ratio equal to 1.0 based on the F606W filter band. The four images are as follows: (i) the original image (upper left), (ii) the host galaxy after removing the point source (upper right), (iii) the best fit model (lower left), and (iv) the residual after subtraction of the best fit model from the original image. The scaling is the same in all images. On the right (upper panel) we show the surface brightness profiles of the various components (i.e. the original data - filled gray circle, the nucleus model/PSF red dotted line, the host galaxy after removal of the PSF (green pentagon), the best fit single sersic model - dashed line and the overall fit as a solid line)  In the lower panel we show the residual after subtraction of the best fit model profile from the data.  
The single Sersic fit indicates a early type galaxy; therefore, no further decomposition into bulge and disk components is necessary. {\bf Middle panel:} AGN ID-417 (B/T=0.3) in the ACS/F850LP filter band.  {\bf Bottom panel:} AGN ID-333 (B/T=0.35) in ACS/F850LP filter band. 
Due to the large contribution of the companion to the surface brightness profile, we model the companion separately and show its component and contribution to the overall
surface brightness profile. Both ID-417 and ID-333 have host galaxy emission that can be decomposed into bulge and disk components.}
\label{q417_decomp}
\end{center}
\end{figure*}

\begin{figure}
\begin{center}
 \includegraphics[clip,angle=0,width=220pt]{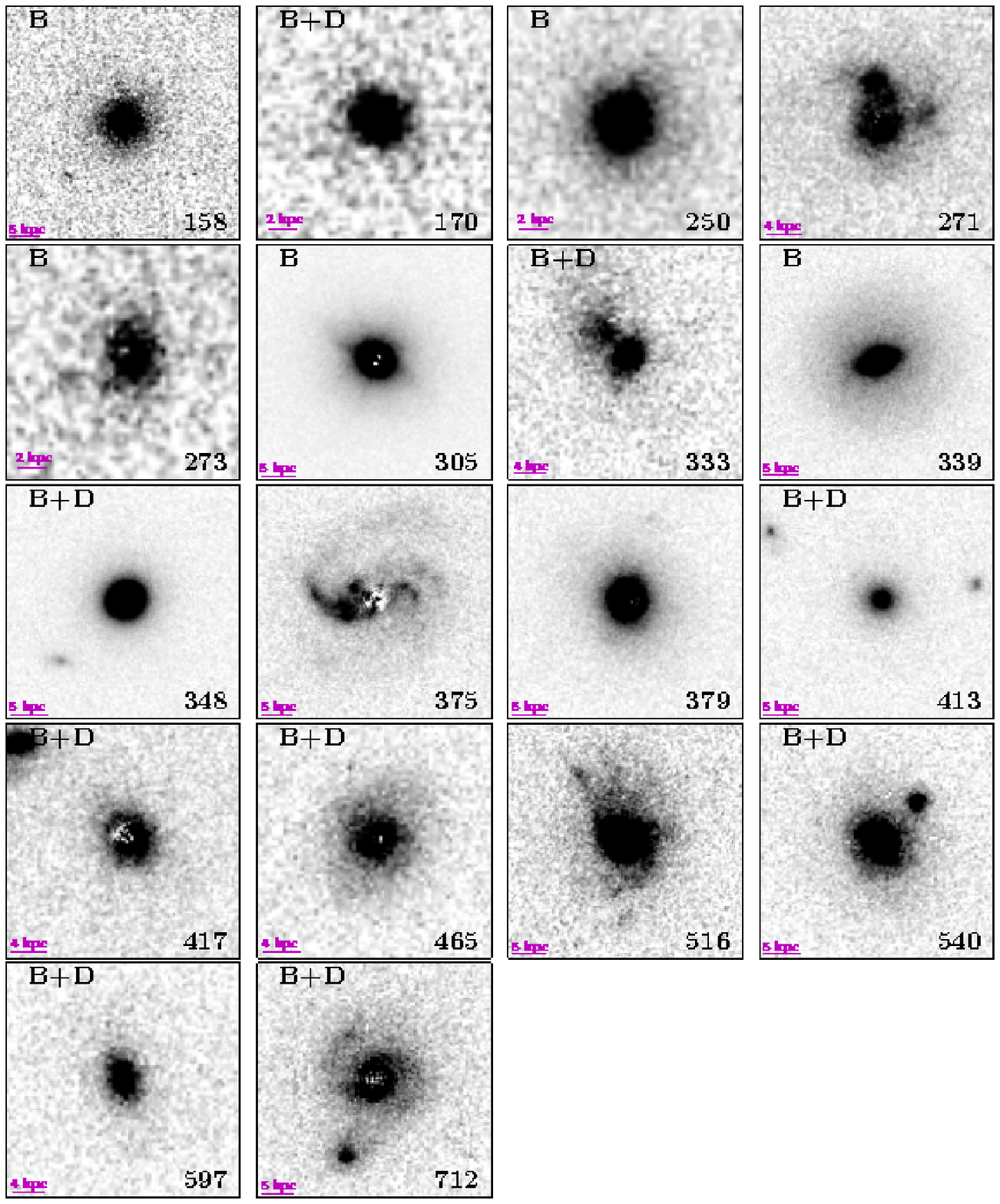}
 \caption{Host galaxy images after removing the point source using GALFIT in the F850LP filter band for the whole sample. We have marked objects for which we could get a direct bulge estimate through either a pure bulge (B) or bulge+disk fit (B+D)}
\label{host_ima}
\end{center}
\end{figure}

 We use GALFIT \citep{Peng(2002),Peng(2010)} to fit the two-dimensional light  
 distribution of each AGN with a point source model represented by an  
 empirical PSF plus a Sersic model for the host galaxy. The nucleus  
 component is either an average PSF created from various field stars  
 around the target, or a single PSF star as described above. The decomposition of the  
 images is done in several steps. First, we conservatively subtract a  
 PSF scaled to the flux contained in a small aperture (typically 2  
 pixels) around the central pixel. For the second step, we perform a  
 full decomposition by adding a Sersic model as a second component. We  
 then optimize the fit to minimize the residuals. This step requires several iterations of the fit
 using different starting values to ensure convergence to a global minimum
in the parameter space. If necessary, we add further components to fit asymmetries in the host (e.g., arm structures).   
 Neighboring galaxies are either masked out or fitted
 simultaneously (see ID-333 for such an example in the bottom panel of Figure 2) to avoid flux spilling over from  
 one object to the other. If the host galaxy flux is below 2\%, we
 decide that the host galaxy is unresolved. 

In Figure \ref{q417_decomp}, we show three representative examples of the image decomposition procedure for objects with different nuclear-to-host (N/H)\footnote{The nuclear-to-host ratio (N/H) is simply the flux attributed to the AGN (N) divided by that of the host galaxy (H).  Both determined through decomposition of the HST images.} ratio and bulge-to-total (B/T) stellar light ratio.  As shown, these cases demonstrate the effectiveness of both the short (F606W) and long (F850LP) wavelength HST imaging.  The top panel shows ID-158 that exhibits a nuclear component attributed to the AGN and a clearly extended component characterized by a Sersic index of 4.2.  With no discernable disk, the morphology is determined to be that of an early-type galaxy.  In the middle panels, the host galaxy of AGN (ID-417) is well resolved above our detection limit even though it has a high N/H ratio.  The Sersic index is found to be n=2.1 that does not favor either a simple early or late type morphology.  Only through a decomposition of the bulge and disk components (as described below) can we determine whether this object is truly bulge or disk dominated.  The third example (ID-333) shown in the bottom panels has a galaxy contribution with a Sersic index of 1.25 indicating a strong disk contribution to the overall morphology.  In Table \ref{T:obs}, we list the sample properties and the results of our fitting routine. In Figure \ref{host_ima}, we show the PSF-subtracted host images of the entire sample.
 
We estimate the uncertainties of our measurements through a series of  
simulations. We create artificial AGN images using empirical PSFs and  
host galaxy models superimposed with artificial noise to match the  
flux levels measured in the real images. We estimate statistical  
errors on the host galaxy and nuclear magnitude by comparing the input  
and output values for our fit parameters. Host galaxy apparent  
magnitude, radius and morphology are extracted from the Sersic model  
fits. Since the Sersic index can be underestimated, in some cases, using GALFIT, especially when the nuclear-to-host (N/H) ratio is high \citep[see][]{Sanchez(2004),Kim(2008a),Kim(2008b)}, we use the simulations to correct  
the Sersic index. Here, we also want to point out the importance of  
the HST data again specially in cases such as ID-333 (See Fig. 2)  
where the AGN is strongly contaminated by a nearby companion that is  
hardly resolved in ground-based data.

A direct bulge/disk decomposition is only possible for objects which have low N/H ratios (typically $N/H<2$).  We find 5/18 host galaxies to have $n_{sersic, corrected} > 3$;  therefore, we classify them as truly bulge dominated.  If the single Sersic fit of the host galaxy indicates the possible presence of a disk  component with $n_{sersic, corrected} <  3$ (13/18 objects), we refit the galaxy with two Sersic models each representing the disk and the bulge.  We put limits on the Sersic Index ($0.5 < n <  1.5$ for the disk and $3 < n < 5$ for
the bulge) of each component to achieve an effectively reduced chi  
square of the residuals as compared to using single values typical  
for a disk (n=1) and bulge (n=4) components. In the end, we are able to directly decompose 14/18 host galaxies 
in the F850LP filter into either a purely bulge or bulge+disk component. Some bulge+disk component fits failed due to the disturbed morphology of the host galaxy (i.e. ID-271 or ID-516) even though the AGN was weak ($N/H<1$). 
Since the F850LP filter provides the best contrast between nuclear and host component, we can use the best fit parameters (i.e. disk/bulge radius, position angle, axis ratios) as constraint in the 
bluer filter bands with typically higher $N/H$ ratios. The best fit radii for the bulge components range from 0.6-6 kpc and are consistent with the typical sizes of elliptical galaxies at similar redshifts \citep{Trujillo(2007)}.
In four cases, the radius of the bulge component is less than 3.5 pixel and the fit can be treated as an upper limit. But we also want to caution that the radii are more sensitive as a free parameter of the fit than the fluxes of the components. 
 
We can further check whether the inclusion of a pseudo-bulge component provides a reasonable fit to the surface brightness profile.  
To do so, we allow the bulge component to have a lower Sersic index.  While it is challenging to distinguish between a pseudo-bulge and a classical bulge with the data in hand, 
we do find stronger residuals in the nuclear region if we fit with a pseudo-bulge component.  These new fits lead to only small changes in the fluxes of each component (<0.1 mag) and bulge-to-disk ratio (<13\%), 
hence only a small impact on the stellar mass estimates.  The poorer fits, seen when incorporating a pseudo-bulge component, may indicate that SMBHs are more directly related to classical bulges than pseudo-bulges \citep{Kormendy(2011)}.

\begin{figure}
\begin{center}
 \includegraphics[clip,angle=0,width=245pt]{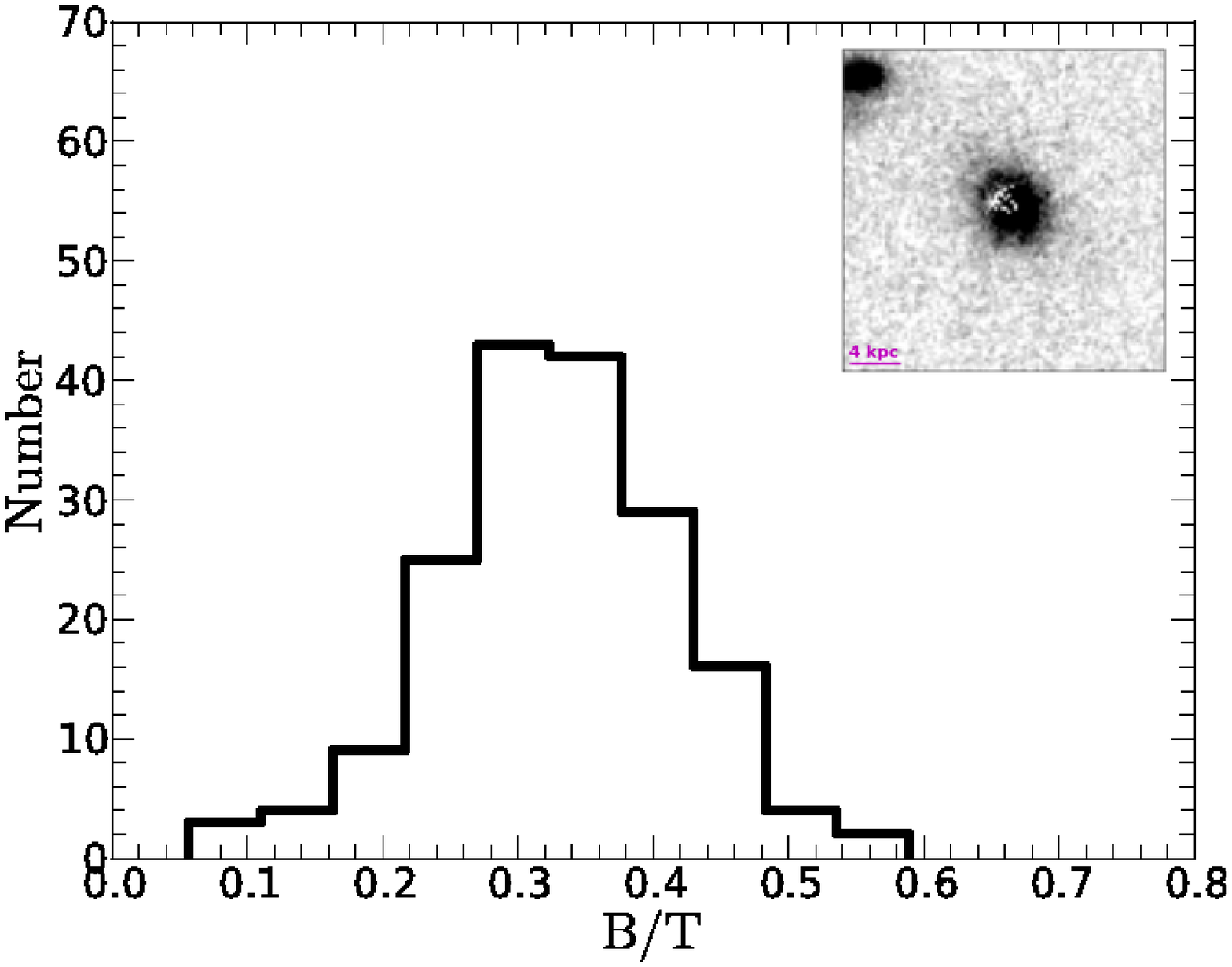}
 \includegraphics[clip,angle=0,width=245pt]{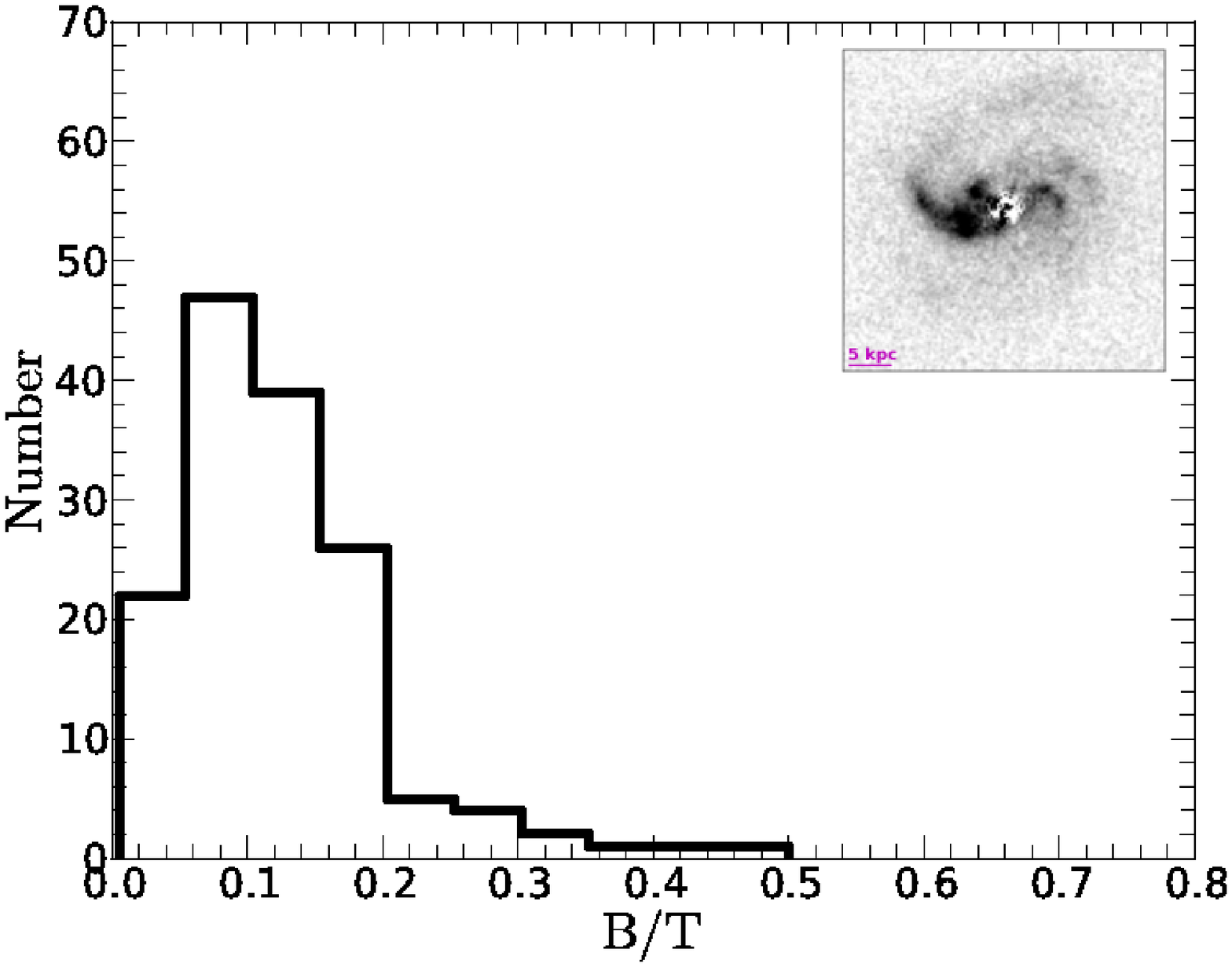}
 \caption{B/T distribution of ID-417 (top) and ID-375 (bottom) extracted from our simulations matching the properties
of the host galaxy using a single Sersic component. The insets shows the host galaxy in F850LP after removal
of the nuclear point source.}
\label{B2T_distr}
\end{center}
\end{figure}

For N/H$>$3, we estimate a bulge correction through extensive simulations rather than a direct measurement.  
For each filter band starting with the longest wavelength due to lower contrast between AGN and host galaxy,
we create a set of artificial AGN+host images using the best fit parameters for the nucleus component. The host  
galaxy consists of two Sersic models, one for the disk and one for the  
bulge component. We vary the free parameters to mimic a broad range of bulges and disks. 
We add appropriate noise measured from the HST images and fit the artificial images with a single Sersic model plus PSF  
model. We compare the fit solutions of the simulations with our single Sersic fit of the real data
and select the B/T ratios of all model fits that recover the original fit parameters within  
their uncertainties. We fit all filter bands separately to account for possible color gradients between the disk and the bulge.
Due to higher N/H ratios at shorter wavelength and typically lower S/N (fainter objects) we constrain some of the free parameters such as size, ellipticity and 
centroid position to the range of solutions found in the F850LP filter band.
In Figure \ref{B2T_distr}, we show the B/T distributions recovered for two of our host galaxies (ID-417 and ID-375) together with an image of the
host galaxy. Figure \ref{B2T_distr} also shows the good agreement between the recovered B/T ratio and the host morphology.
For example ID-375 has a low B/T ratio of about 0.1 indicating the presence of a dominating disk and the image of the host galaxy shows some
spiral structures with very little light concentration in the center of the galaxy.
The mean value of the B/T distribution and its uncertainty is then used to estimate the bulge mass. The error bars on the photometry  
are typically larger by 0.15-0.3 mag.

\subsection{Total/Bulge Stellar Mass Estimates}

Our main goal is the estimation of the stellar mass content of each  
host galaxy and its bulge component in the sample by
converting the rest frame optical colors into mass-to-light ratios.  
For targets in the GEMS area, we have only a single optical
color while for GOODS we have multiple colors based on four filter  
bands. This method has been successfully employed in several studies  
on AGN host galaxies \citep[e.g.,][]{Schramm(2008),Jahnke(2004),Jahnke(2009),  
Sanchez(2004)}. As demonstrated by Bell et al. (2003)  
for a variety of star formation histories, the stellar mass-to-light ratio (M/L)  can be  
robustly predicted from the B-V  color. We adopt their formula  
based on a Chabrier IMF:

\begin{equation}
\mathrm{log}_{10}(M/L_V)=-0.728+(1.305\times (B-V))\:,
\label{Eq_stell_mass}
\end{equation}
where $M/L_V$ is given in solar units. The choice of an IMF has a systematic effect on the final mass  
estimation. Using a Salpeter IMF would typically increase our mass-to-light ratios by factor of   1.4.

The ECDFS area is well covered by broad-band photometry from various   
instruments ranging from the ultraviolet to the infrared.  The available  
photometry provides another approach to estimate the total stellar  
mass content of the host galaxies through direct SED fitting (as shown  
by Merloni et al. 2010). Although this would be a powerful  
alternative, the data suffers from additional uncertainties such as  
source confusion in the ground based data or variability of the  
sources due to multi-epoch data. In any case, we decided to implement  
SED fitting only as a consistency check on our total stellar mass  
estimates based on the HST data. For the procedure, we use our own  
algorithm based on a Levenberg-Marquart $\chi^2$ minimization. We use a set of SED model templates \citep{Maraston(2005)} with declining star formation histories based on a Kroupa IMF (which gives similar results as a Chabrier IMF), solar metallicity and a dust extinction law following \citet{Calzetti(2000)}. We make use of the broad-band image  
decomposition (AGN+host) based on the HST results to constrain the template  
models that includes the photometric errors in each filter band.  
First, we fit a template AGN model (Richards et al. 2006) to the  
photometry of the nucleus obtained from the decomposition and subtract  
this from the total (ground-and/or space-based) photometry. Next, we fit the residual fluxes with  
either a single template or a two component template.  Strong  
contamination from unresolved sources in the ground based data (i.e.  
in ID-333) are taken into account and subtracted separately using the  
HST photometry as an additional constraint for the companion template  
model. To estimate errors on the
stellar mass, we use a Monte Carlo approach. We vary the observed  
flux in each bandpass by a random number which is
Gaussian distributed with a sigma defined by the flux error. We generate  
100 simulated SEDs and recompute the fit.  Masses
from both approaches typically agree within 0.13 dex. In Figure 6, we  
show four examples (ID-339, ID-170, ID-465 and ID-250) of the SED decomposition compared to 
the single epoch spectrum used for the BH mass estimation. 
These four AGN represent different level of AGN and host stellar continuum
throughout our sample. In addition, these objects also have  
different B/T ratios: 1.0 (ID-339, ID-250), 0.5 (ID-170) and 0.24
(ID-465).  These examples illustrate the clear advantage gained from having the HST  
photometry. Only with HST resolution, we can constrain the flux of the AGN and the host galaxy including the bulge and disk components separately. The mass estimates from \citet{Bell(2003)} and SED 
fitting typically agree within 0.15 dex.

In Figure \ref{color_mass}, we plot the U-V rest-frame color versus the total stellar mass. Nearly
all hosts concentrate at the high mass end and below the red sequence (i.e., the green valley). The masses of our host galaxies are comparable to typical red sequence galaxies but
the colors of the host indicate a population of recently formed stars.

\begin{figure}
\begin{center}
\includegraphics[clip,angle=0,width=200pt]{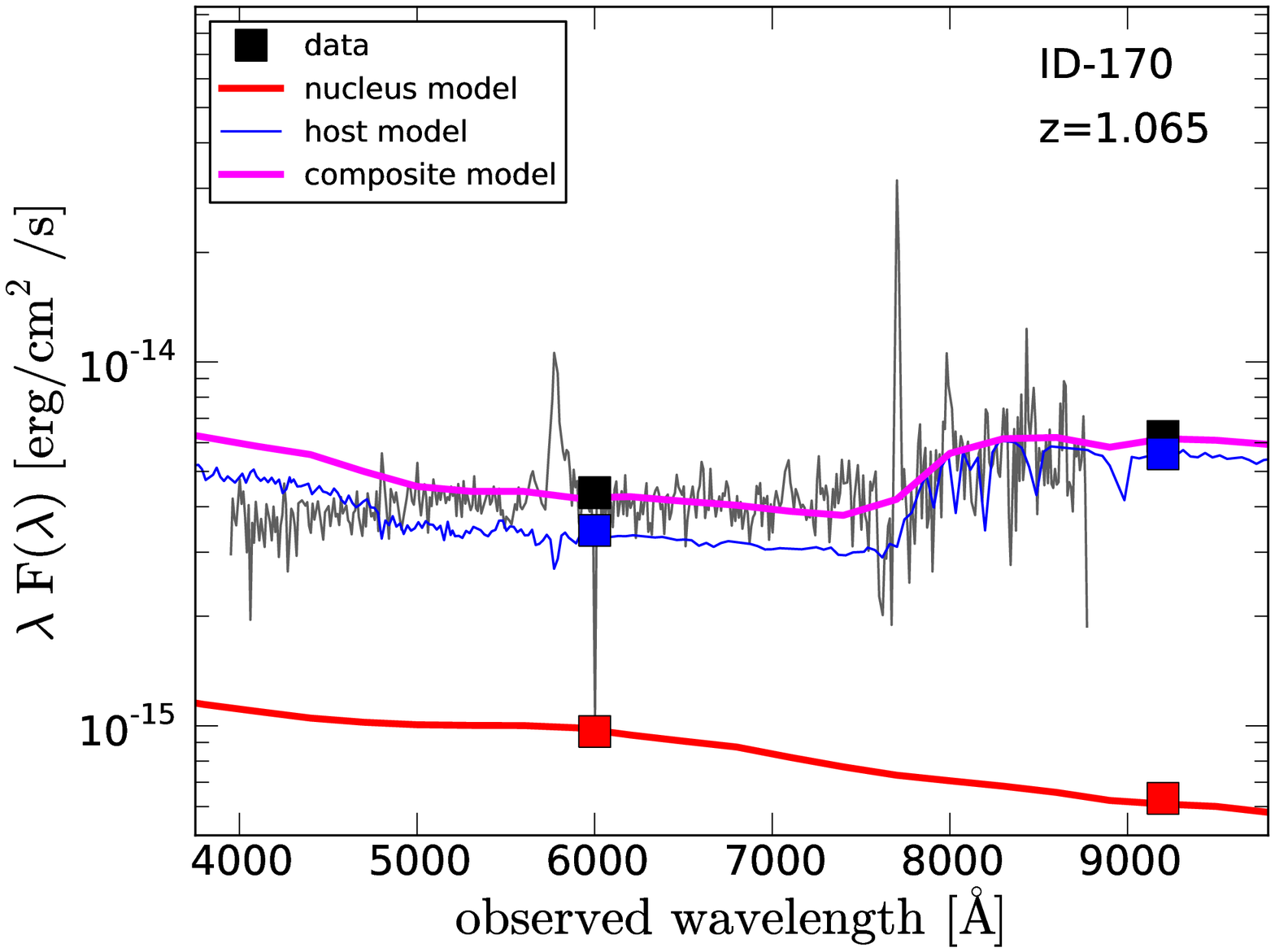}
\includegraphics[clip,angle=0,width=200pt]{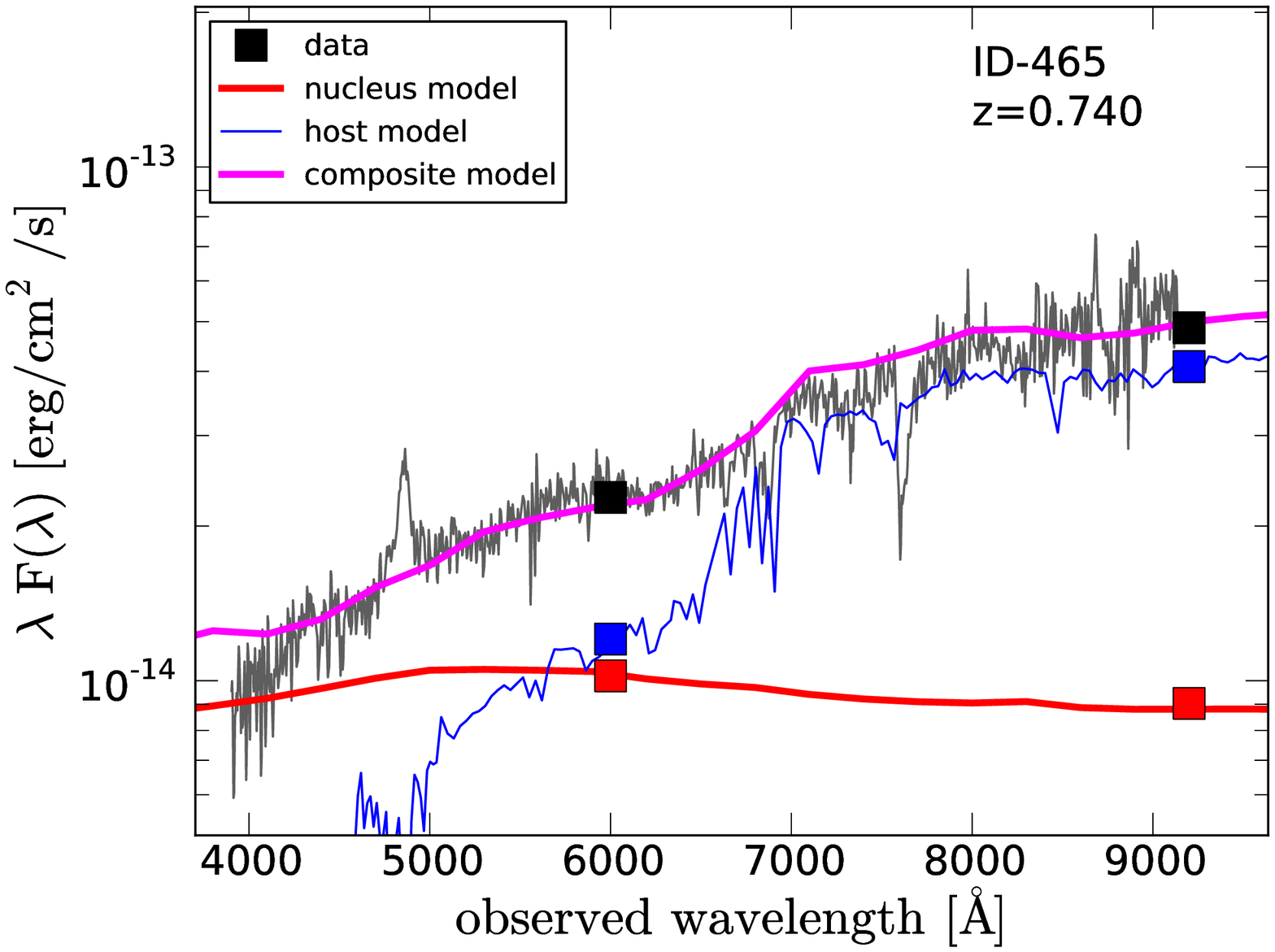}
\includegraphics[clip,angle=0,width=200pt]{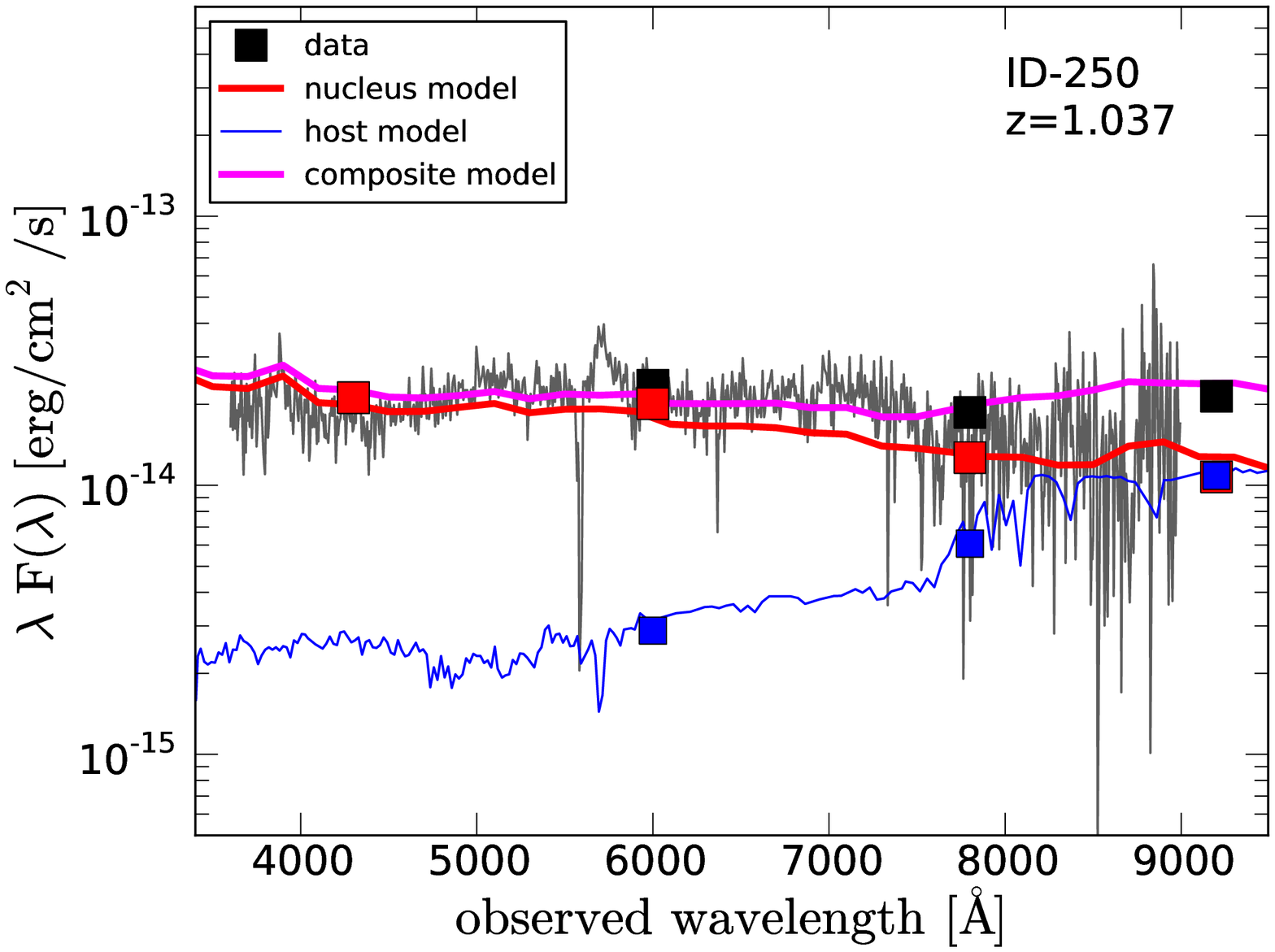}
\includegraphics[clip,angle=0,width=200pt]{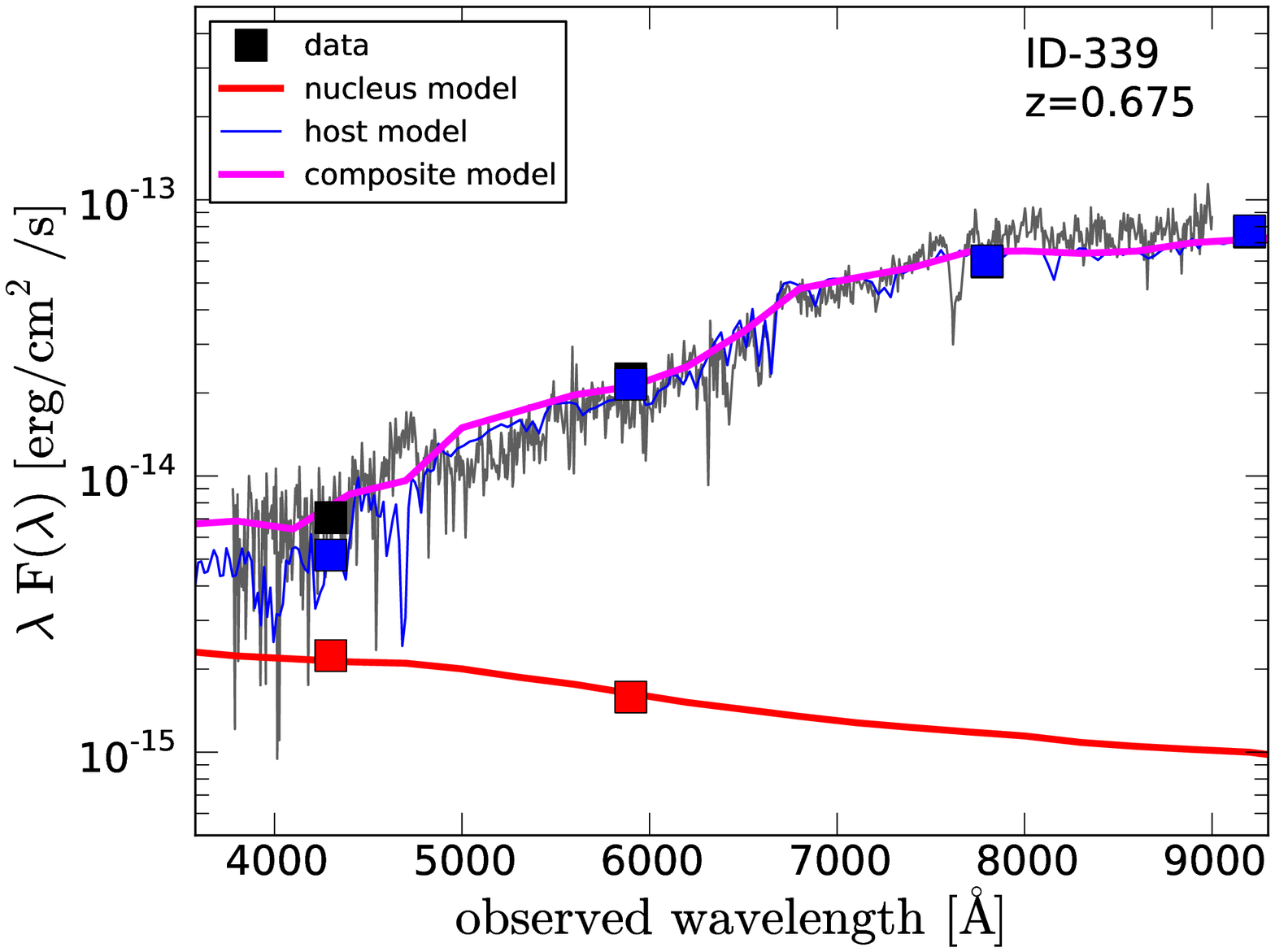}

 \caption{Examples of the SED template fitting for different nuclear-to-host ratios (ID-339,ID-250; ID-170; ID-465).  Black squares show the total photometry measured in the bands covered by HST (either F606W and F850LP for GEMS or F435W, F606W, F775W, F850LP for GOODS).  Red and blue squares show the results from the image decomposition in the HST filter bands representing the nuclear component (red) and the host component (blue).   The red and blue solid lines represent the best fit AGN and host galaxy models with the magenta solid line showing their sum.}
\label{q339_sed}
\end{center}
\end{figure}

\begin{figure}
\begin{center}
 \includegraphics[clip,angle=0,width=245pt]{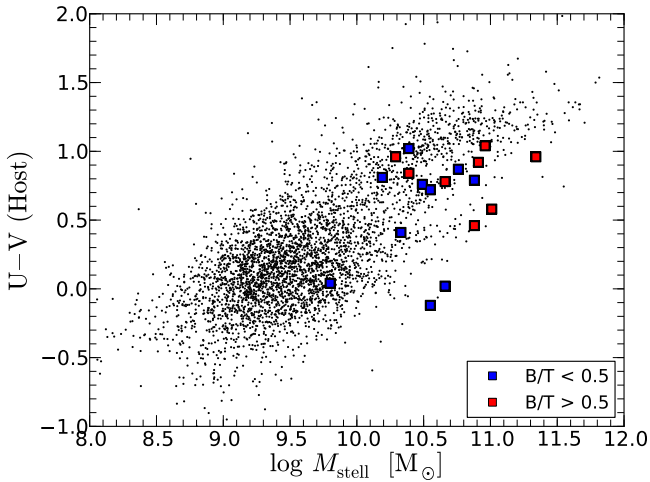}
 \caption{Rest frame U-V color vs. total stellar mass of the host galaxy compared to a sample of inactive galaxies in the same redshift range taken from the GEMS catalog.}
\label{color_mass}
\end{center}
\end{figure}

\section{Black Hole Masses}

  We measure black hole masses for our entire type 1 AGN sample using  
  single-epoch spectra that provide both a velocity width of a broad  
  emission line and the monochromatic luminosity of the continuum.
  We use optical spectra acquired mainly from the followup of X-ray sources \citep{Szokoly(2004),Silverman(2010)}.  We supplement these with spectra taken with FORS2 on the VLT but not yet publicly available.
  
  Several prescriptions to estimate black hole masses are available from  
  the literature using various emission lines such as H$\beta$, MgII or CIV  
  \citep{Kaspi(2000),Vestergaard(2006),Collin(2006), McLure(2002)}.  
  Due to the redshift range of our sample  
  and optical spectroscopic coverage, we use the MgII emission line to  
  estimate virial black hole masses in all cases. Although most of
the black hole mass calibrations are based on reverberation mapping
 data of $H\beta$ several studies have shown that there is good agreement between 
the mass estimates based on MgII and the Balmer lines ($H\beta$, H$\alpha$) out to high redshifts (Shen \& Liu 2012; Matsuoka et al. 2012) by combining optical and NIR spectroscopy. The prescription for  
  estimating black hole mass as given in \citet{McLure(2002)} is  
  implemented; although, we recognize that similar recipes are available  
  elsewhere (Kong et al. 2006, McGill et al. 2008, Wang et al. 2009)  
  with each of these agreeing essentially to within 0.2-0.3 dex.
  
  \begin{figure*}
\begin{center}
\includegraphics[clip,angle=0,width=400pt]{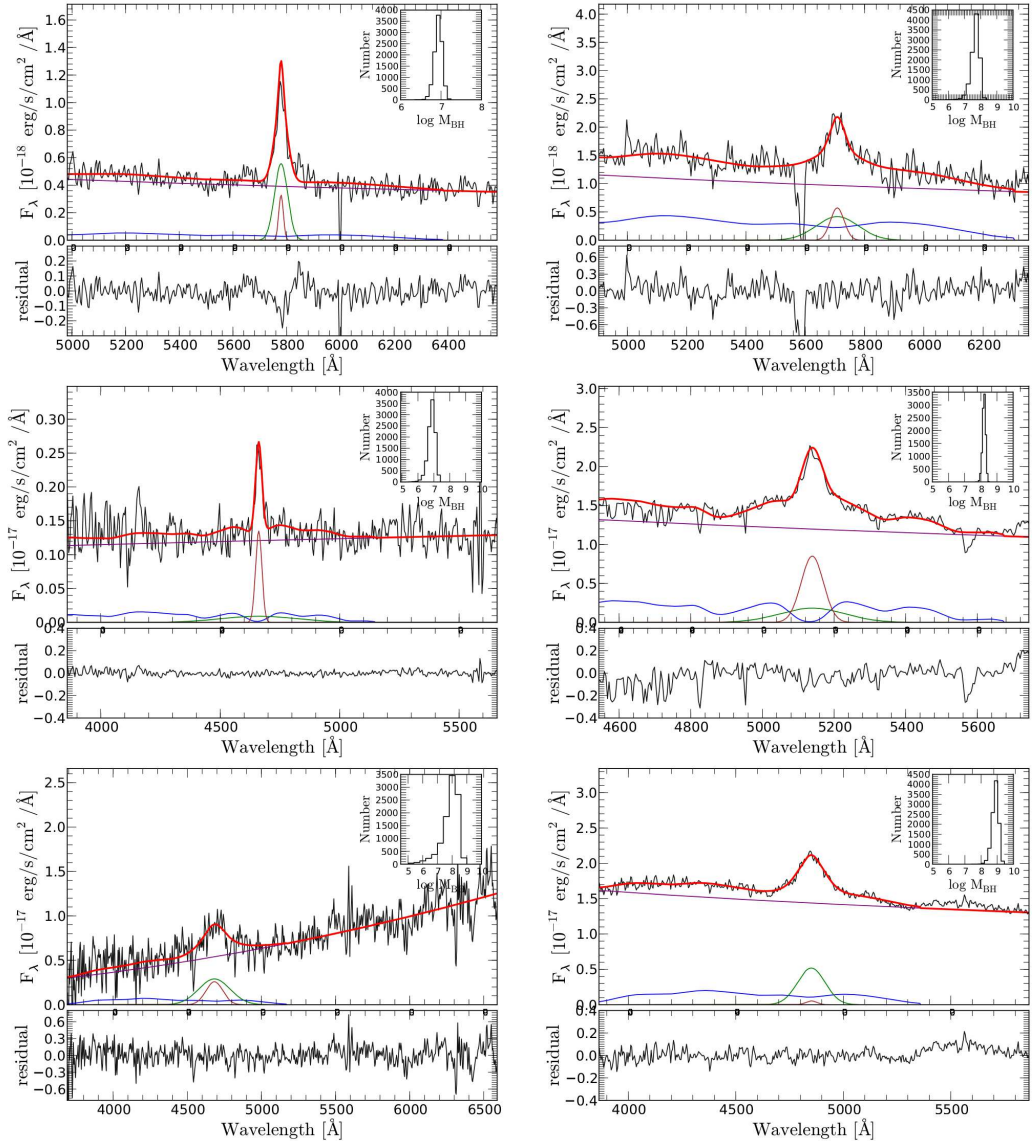}

 \caption{Multicomponent fit to broad MgII emission line for six representative objects from our sample. The top panel shows the spectral range around the emission line. The best fit model is indicated as a red solid line. The different components are as indicated:
Fe-emission (blue), pseudo-continuum (purple), Gaussian components (green and brown). The residual (data-fit) is shown in the lower panel. In the upper right corner of each panel, we show the black hole mass distribution computed from our 
Monte Carlo tests based on the uncertainties of the line width and continuum luminosity measurements. The objects are (from top left to bottom right): ID-170,ID-250,ID-413,ID-417,ID-339, and ID-379}
\label{specfit}
\end{center}
\end{figure*}

  We perform an iterative least-squares minimization to fit the MgII line for each AGN to measure its line width. Our procedure is a modified version of the one used in \citet{Gavignaud(2008)}. 
  The number of components to fit the line depends on the characteristics of the objects and quality of the data. 
  We fit the region around the emission line using a model that includes a pseudo-continuum and one or two Gaussian components to characterize the line profile.
  We find that for the local continuum a powerlaw+broadened Fe-template \citep[provided by M. Vestergaard: see][]{Vestergaard(2001)} gives the best results. 
  Specially the strength of the Fe-emission in the wings of the MgII line can vary strongly (see ID-250 for strong Fe and ID-170 for very weak Fe) and affects the outcome of the fit.  We try to both minimize the number of model components and optimize the residuals around the emission line.
  We either interpolate over absorption features or mask them out.  A FWHM of the line profile is determined using either a one or two component Gaussian model. 
  We have tested the same algorithm on the sample from \cite{Merloni(2010)}. Even tough we find some scatter for the individual fits, there is no systematic offset in the final black hole mass estimates. Two of our objects overlap 
  with the study from \cite{Bennert(2011)}; our mass estimates agree within ~0.1 dex using the same recipe.

  In the next step, we measure the continuum luminosity at 3000 \AA\, required to estimate a radius  
  to the BLR. For luminous AGN ($L_\mathrm{bol}>45$) the continuum luminosity can be directly measured
  from the spectrum due to the typically low impact of the host galaxy. For our sample, we find that in several cases, 
  there is a significant host galaxy contribution that must be taken into account (see Fig. \ref{q339_sed}). Therefore, we decided
  to measure the monochromatic luminosity at 3000 \AA\, by decomposing the HST/ACS images. The  
  procedure enables us to isolate the AGN (i.e., nuclear) emission from  
  its host galaxy most effectively. We then fit an average quasar SED  
  template \citep{Richards(2006)}, accounting for dust attenuation to  
  estimate the intrinsic continuum luminosity at 3000 \AA. We find that the continuum luminosity based on HST imaging agrees with that determined from the decomposition of the broad-band SED to within 5\%. Monte Carlo realizations using the uncertainties of the FWHM and $L_{3000}$  
  measurements enable us to estimate the uncertainties on the black hole mass  
  in addition to 0.4 dex uncertainty inherent in the scaling relations. In Figure \ref{specfit}, we present examples of the fits to the broad emission lines in six AGN with different quality of data.  
A summary of the results of our line fits are shown in Table 1.

\section{Results}

\subsection{The BH Mass-Total Stellar Mass Relation }

\begin{figure*}
\begin{center}
  \includegraphics[clip,angle=0,width=240pt]{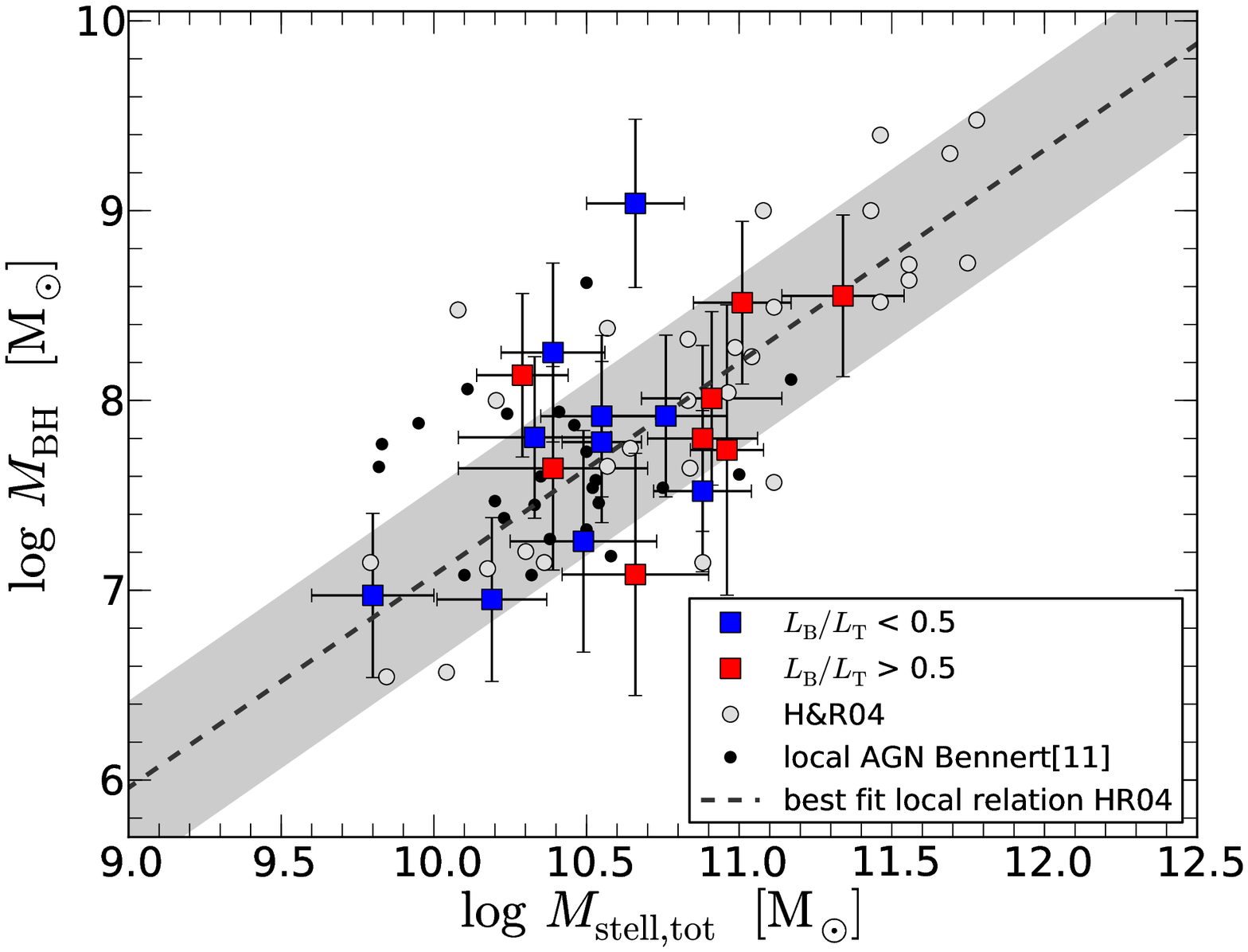}
 \includegraphics[clip,angle=0,width=240pt]{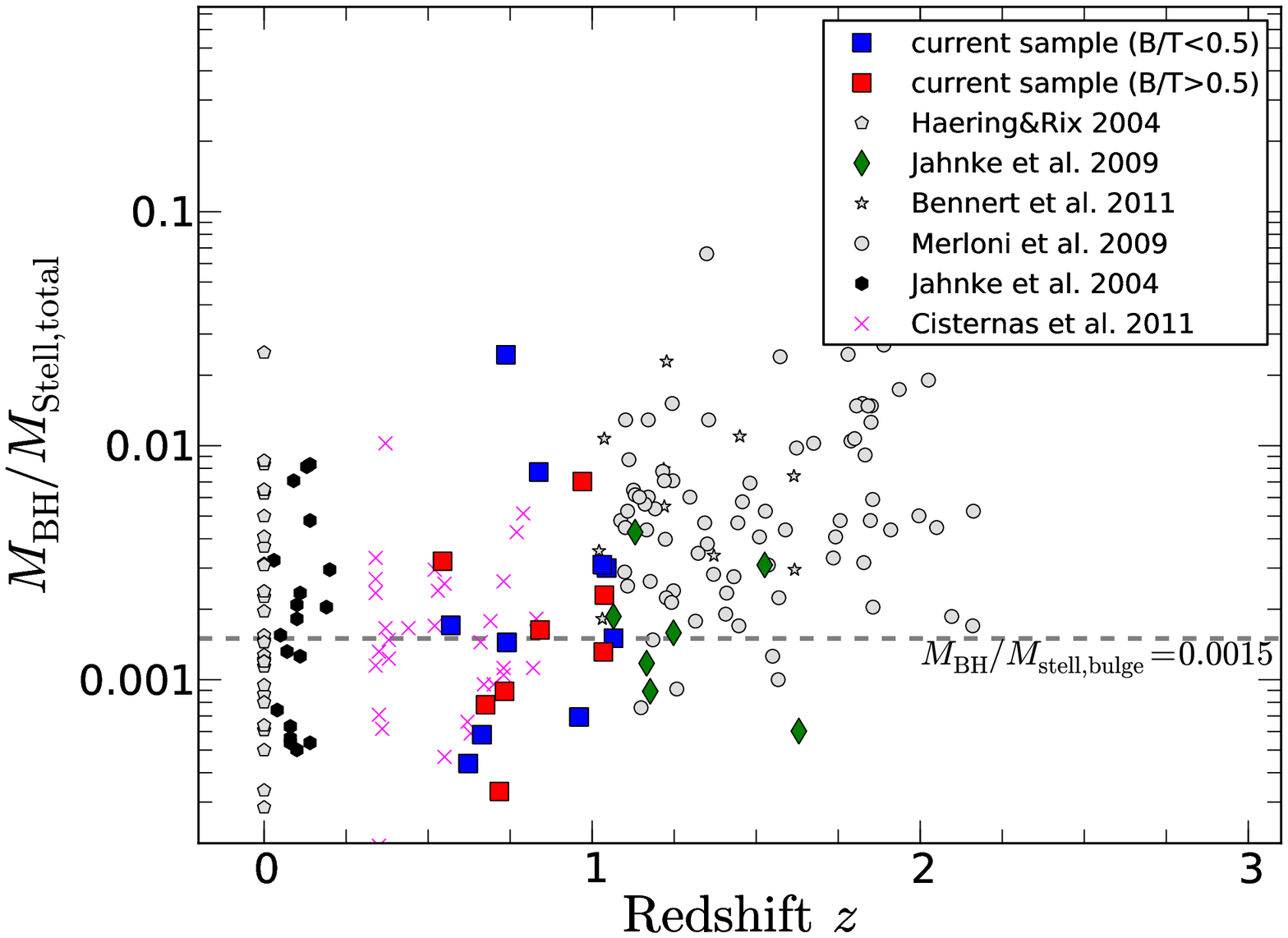}
 \caption{{\it Left:} \mbhtot relation for our sample of intermediate redshift AGN. The sample has been divided into bulge (red squares) and disk dominated (blue squares) 
systems based on our bulge corrections. The data is superimposed on the sample of inactive galaxies from H\"{a}ring \& Rix (2004) (dynamical mass) and the local AGN (stellar mass) sample from \cite{Bennert(2011a)}. The dashed line represents the best 
fit from H\"{a}ring \& Rix 2004 with a 0.3 dex scatter shown as the grey shaded area. {\it Right:} Redshift evolution of the 
relation in comparision with several other studies taken from the literature. The dashed line shows the mean constant ratio from the local relation by H\"{a}ring \& Rix (2004).} 
\label{mbhmtot}
\end{center}
\end{figure*}

We first present the relation between black hole mass ($M_\mathrm{BH}$) and total stellar mass ($M_\mathrm{*,Total}$) in Figure~\ref{mbhmtot} ($left$ panel).  From the distribution of data points, it is apparent that our sample does not have the dynamic range in either stellar mass or black hole mass to establish both a slope and normalization simultaneously of a linear fit.  Fortunately, we can compare with the local relation established using inactive galaxies (mainly ellipticals or S0) as done by \citet{Haering(2004)} and determine whether an offset exists.  We find that essentially all of our AGN fall along the local \ratio relation.  It is important to highlight that the bulge mass is equivalent to the total stellar mass for the local comparison sample.  To be more specific, we find that 17/18 objects, considering their 1$\sigma$ errors, are consistent with the typical region of 0.3 dex scatter around the best fit local relation having a slope of 1.12 \citep{Haering(2004)}.  Given our limitations in mass coverage as mentioned above, we fit a linear regression model to our data while fixing the slope to the value given above thus determining only the normalization. We find the best-fit normalization to be 8.31 by using FITEXY \citep{Press93}, which estimates the parameters of a linear fit while considering errors on both variables. The fit is affected by the single target offset from the relation. If excluded for no obvious reason, the constant would be 8.24.  With a simple Monte Carlo test, we can reject the null hypothesis that the two samples are significantly different. While the local inactive sample is established using dynamical masses, we do not expect these to differ substantially from the stellar masses; this is in fact the case as demonstrated in \citet{Bennert(2011a)}.

Ideally, we would like to compare our sample with a local sample of active SMBHs with stellar mass measurements of their hosts. The work of \cite{Bennert(2011a)} allows such a direct comparison.  We show these data in Figure~\ref{mbhmtot} as marked by small black circles. 
Carrying out the same fit as for the ECDFS AGNs, we find the best fit constant to be 8.30 for the local AGNs. We use the total stellar mass for the regression fit of both active samples (\cite{Bennert(2011a)}, ECDFS AGNs) and find no significant deviation between them in the \mbhtot relation.

Our result agrees well with findings of recent studies of the $M_\mathrm{BH}-M_\mathrm{*,Total}$ relation at high redshift.  In particular,  \cite{Jahnke(2009)} 
use a similar technique of decomposing HST images and converting rest-frame optical colors into stellar mass-to-light ratios based on a sample of AGNs at $z>1$ in COSMOS with NICMOS coverage.  Their sample consists of ten objects with seven for which they achieve a decomposition in multiple bands and find no offset in 
black hole mass given their total stellar masses. Our study effectively improves the statistics by a factor of 2.5 and fills in a gap in redshift coverage (see Figure~\ref{mbhmtot}, $right$ panel).  In addition, these results are supported by the findings of \citet{Cisternas(2011)} who explored the same relation on a sample of BLAGN at $0.3<z<0.9$ from the COSMOS survey; although, only one HST band is available to constrain the stellar mass content of the host galaxy. 

Taken together, these studies \citep{Jahnke(2009),Cisternas(2011)}, including our own, clearly contrast with other works at high redshift that claim an increasing offset in black hole mass for a given stellar mass.
In the right panel of Figure \ref{mbhmtot}, we show the redshift evolution of $M_\mathrm{BH}$-$M_\mathrm{*,Total}$ ratio compared to various other studies probing the same relation.
Some studies are using different mass estimators for the black hole masses or stellar masses. For example, \cite{Merloni(2010)} use the prescription of \cite{McGill(2008)} for their black hole mass estimations and assume 
a Salpeter IMF for their stellar mass estimates. When necessary, we convert the masses of different studies to the prescription based on the formula from \cite{McLure(2004)} and a Chabrier IMF for the stellar mass estimates. In case of \cite{Merloni(2010)},
the corrections have only a marginal effect on the $M_\mathrm{BH}-M_\mathrm{*,Total}$ relation.
Based on our results, we cannot confirm or rule out a stronger evolution at higher redshift ($z>1.5$). In particular, our mean bolometric luminosity is log~$L_{bol}=44.7$ while the higher redshift sample of \cite{Merloni(2010)} is at $log~L_{bol}=45.5$. As a consequence, the mean BH mass is shifted to higher masses and 
therefore a direct comparison with these objects and any trend implied by the data might be biased by the differences in the sample properties.  It is worth highlighting that our results are likely to be less biased due to selection since our BH masses are typically below $10^9M_\odot$, the knee in the black hole mass function; we may be effectively avoiding the problems fully presented in \citet{Lauer(2007)}.

\subsection{The BH Mass-Bulge Stellar Mass Relation }

While the total stellar mass is well-determined using different methods \citep[][]{Schramm(2008),Jahnke(2009),Merloni(2010),Cisternas(2011),Bennert(2011)}, we usually do not know how much of the total mass is present in the bulge. As
stated above, only 5/18 of our AGN hosts have a Sersic Index $n>3$ indicating a purely bulge-dominated host galaxy. We make the assumption that for these objects the total mass is the same 
as the bulge mass. For the remainder, we estimate the bulge contribution to the total mass by corrections to the total mass by accounting for the contribution of the disk. Applying 
the same cut at $n<3$, we find that $\sim$72\% of the host galaxies show a disk component. Although the fraction is in good agreement with
the results presented by \citet{Schawinski(2011)} on a sample of X-ray selected AGN in the $Chandra$ Deep Field South at $2<z<3$.  Although, we draw a different conclusion
on the importance of the disk component, in terms of the mass contribution to the total mass.  Our bulge/disk decomposition shows that, even though a disk is present, the mass of the central bulge can still dominate the total mass of the host galaxy.  The different redshift regimes might play an important role since there is about 3-5 Gyr of galaxy evolution between our study and that of \citet{Schawinski(2011)}. Using the B/T ratio to divide our sample into bulge and disk dominated systems, we find that $\sim50$\% of the sample has a significant bulge component with $B/T>0.5$; this can even be true for objects with a surface brightness profile of the host galaxy described by a fit with a Sersic index of $\sim$2.

\begin{figure*}
\begin{center}
  \includegraphics[clip,angle=0,width=350pt]{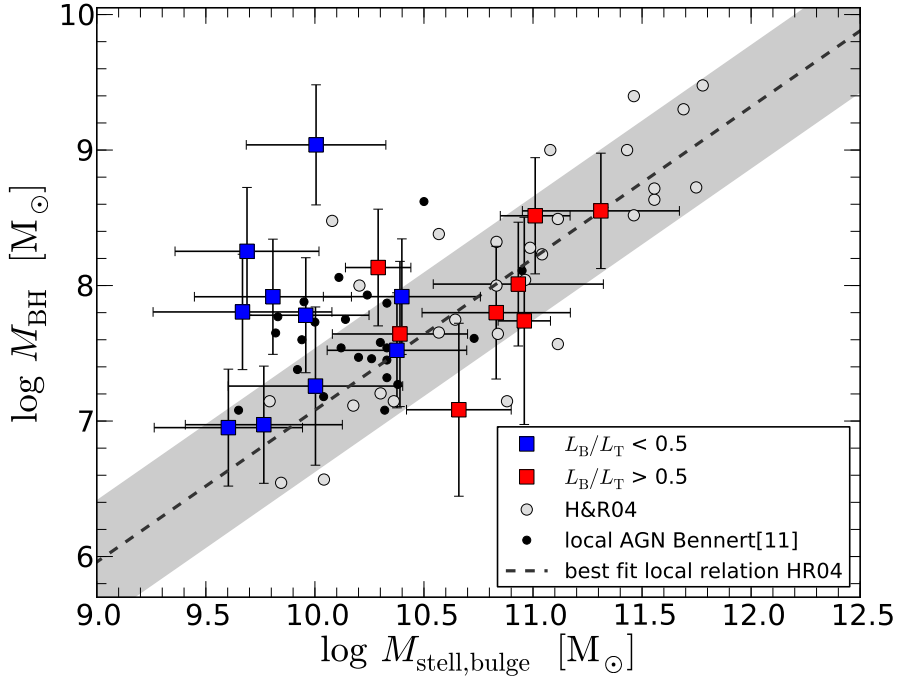}
\caption{\ratio relation for our sample of intermediate redshift AGN. Symbols are the same as in Figure \ref{mbhmtot}
 }
\label{mbhmbul}
\end{center}
\end{figure*}

\begin{figure}
\begin{center}
  \includegraphics[clip,angle=0,width=225pt]{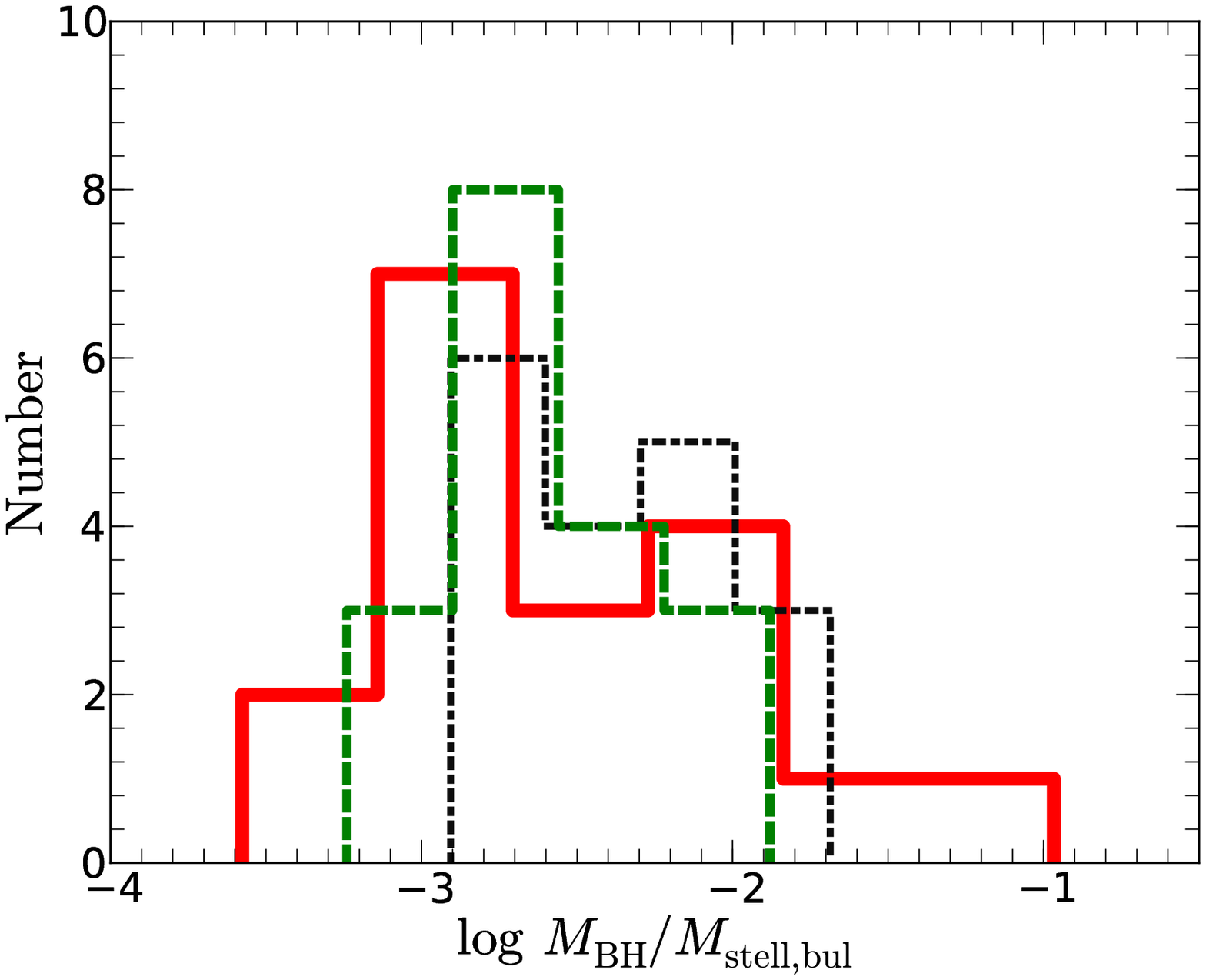}
\includegraphics[clip,angle=0,width=225pt]{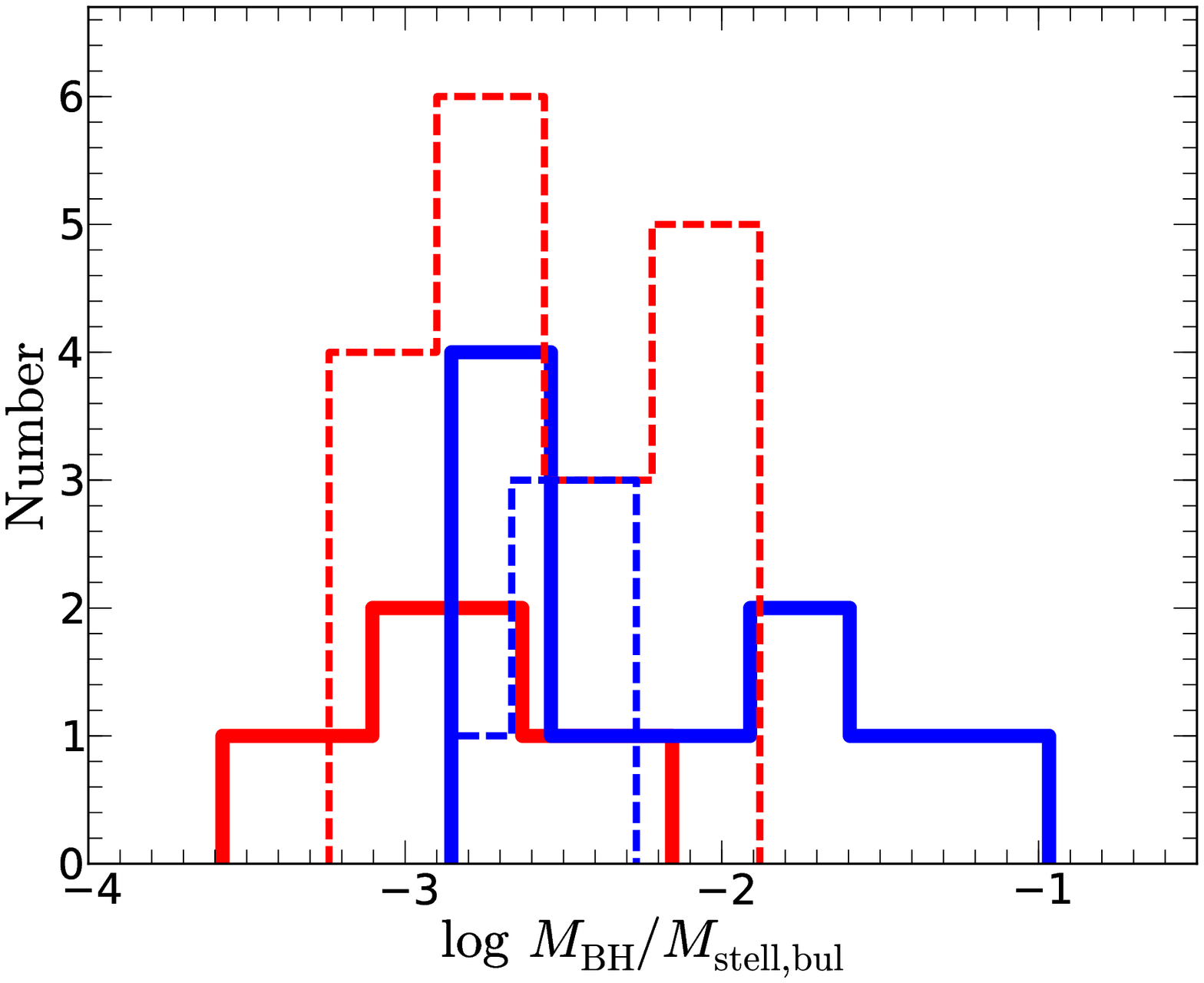}
\caption{$Top$ panel: \ratio ratio distribution for our sample of intermediate redshift AGN (red solid) compared to the sample of inactive (black dotted) and active (green dashed) galaxies from \cite{Bennert(2011a)} using stellar host galaxy masses. 
The inactive sample from \cite{Bennert(2011)} is based on the group 1 sample from \cite{Marconi(2003)} which overlaps also with the sample of H\"{a}ring \& Rix 2004.
$Bottom$ panel: Comparision of the \ratio ratio 
distribution for the bulge dominated ($L_{B}/L_{T}>0.5$) subsample shown as red solid line, the disk dominated ($L_{B}/L_{T}<0.5$) subsample shown as blue solid line.
The data is superimposed on the sample of local AGN from \cite{Bennert(2011a)} showing the sample separation into bulge (red dashed) and disk (blue dashed) dominated systems.
  }
\label{mbhmbul_hist}
\end{center}
\end{figure}

We can now establish the \ratio relation at $0.5<z<1.2$. In Figure \ref{mbhmbul}, we plot the 
\ratio relation and compare our results with the sample of inactive galaxies from H\"{a}ring \& Rix (2004) and local AGN from \cite{Bennert(2011a)}.
The stellar mass measurements for the local AGN allow a more direct comparison with our sample than the dynamical masses of H\"{a}ring \& Rix (2004).  We find that the mass distributions for all three samples are very similar with each other.  This can be clearly seen in a histogram of the mass ratio ($log~M_{BH}/M_{*,Bulge}$) shown in the top panel of Figure \ref{mbhmbul_hist}, where there is no significant difference in the median value.  Overall, we find that 78\% of the AGNs are consistent with the local relation.  If we consider the single object undergoing a clear major merger (ID-333), there are only three objects that are significantly offset from the local relation.  If we artificially move this object onto the relation, then 83\% of AGNs in our sample are consistent with the local relation.  We interpret this as evidence for a black hole-bulge relation, at these redshifts, to be similar to the local relation.  Interestingly, we do find additional scatter in our sample compared to that in the local distributions.  We further note that there are no objects well below the \ratio relation. 

We can further investigate where high-z AGN lie in respect to the local relation as a function of their bulge-to-total ratio. All objects with $B/T>0.5$ fall nicely onto the local relation (see Figure~\ref{mbhmbul} and the bottom panel of ~\ref{mbhmbul_hist}). They are also the most massive objects in the sample in terms of their bulge mass.
Objects with a $B/T<0.5$ are clearly separated in bulge mass (from bulge-dominated objects) and the majority are still in good agreement with the local relation.  Only four
objects have under massive bulges considering their $1\sigma$ error bars including ID-333 which has a massive companion that might move the whole system onto the relation after the merger.  

\section{Discussion}

An important question for SMBHs and their host galaxies is their subsequent evolution in the black hole - bulge mass plane.  As previously mentioned, 83\% of the bulges in our sample are already massive enough that their \ratio ratio agrees well with that seen in inactive galaxies today (see Figure~\ref{mbhmbul}). We illustrate this further in Figure \ref{mbhmbul_hist} (top panel), by comparing the distribution of the \ratio ratio between various samples. Interestingly, there are some outliers with undermassive bulges, relative to their BH mass, that are preferentially disk dominated galaxies. In the bottom panel of Figure \ref{mbhmbul_hist}, we compare the distributions of this ratio for the bulge and disk dominated subsamples separately to the distribution of the local AGNs.  Even though the number statistics are small, we find no difference for the bulge dominated subsample by looking at their median ratios.
The situation is different for objects in the disk-dominated subsample.  While some objects overlap with the distribution \ratio ratios of the local AGN, the median ratio of the
disk dominated subsample is shifted by 0.5 dex towards a higher ratio. When comparing bulges of similar mass ($\mathrm{log}\,M_\mathrm{*,Bulge}<10.5$), local AGN host galaxies have a smaller offset ($\sim$0.25 dex).
On the other hand for the same mass matched subsample which includes 22/25 objects in the local AGN sample and 12/18 from our sample, we find that the local AGN sample contains only $\sim$30\% disk dominated systems while our
subsample contains $\sim80\%$ disk dominated systems. Within the AGN population, we may be witnessing both a migration onto the local relation and a morphological transformation with cosmic time. We recognize that selection 
effects may impact such comparisons. Ideally, we want to have an AGN sample spanning a wide baseline in redshift with equivalent selection, BH mass indicators and sufficient statistics. 

This leads to the question how these host galaxies can grow their stellar bulge mass to match the bulge masses
seen today. One possible track could be the event of a major merger that leads ultimately to a significant increase in stellar bulge mass. Mergers are seen to play a 
role in black hole growth for similar X-ray selected samples \citep[][]{Silverman(2011)}. Out of our 18 AGN, only one (ID-333) shows
signs of an ongoing major merger.  Even though other host galaxies do show some signs of minor merger activity, we conclude that the growth of the bulge through a major merger event
in the near future is not certain. On the other hand, the good agreement of the AGN host galaxy 
$M_\mathrm{BH}$-$M_\mathrm{*,Total}$ relation with the local relation clearly shows that all the mass needed to put our host galaxies onto the 
local \ratio relation is already in place within these galaxies at redshift $z\sim1$ \citep{Jahnke(2009)}. Therefore, mass transfer from the disk to the bulge is neccessary to grow their bulges. 
Any bulge growth through internal processes has to overcome the mass growth of the black hole otherwise the galaxy would just move on a diagonal track in the 
\ratio relation.  

\begin{figure}
\begin{center}
  \includegraphics[clip,angle=0,width=225pt]{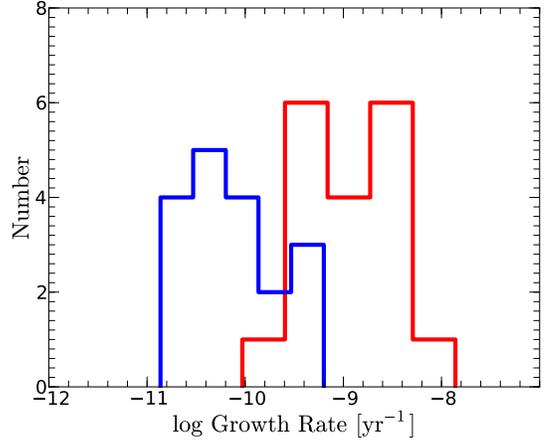}

\caption{Comparision of the growth rate distributions of the host galaxy (blue; $M_{*,Total}/SFR$) estimated from the UV-SFR and the BHs (red; $M_{BH}\times(dM/dt)^{-1}$) of our sample}
\label{growth}
\end{center}
\end{figure}

While the BHs in their active phase are growing, we can also investigate how the host galaxy is growing in stellar mass
by looking at their individual growth rates (i.e., SFR) and compare these to the BH growth rates. We estimate star-formation rates based on the UV continuum 
from our best fit SED models and converted these into growth rates (SFR/$M_\mathrm{stell}$). In Figure~\ref{growth}, we compare the growth rates of the host galaxies with the growth rates 
of the BHs as determined by $\dot M /M_\mathrm{BH}$. $\dot M$ is determined from $L_{bol}=\epsilon \dot M c^2$. To estimate the bolometric luminosities $L_{bol}$ 
and Eddington ratios, we use the luminosity dependent corrections from \cite{Hopkins(2007)} applied to our derived continuum luminosities at 3000 \AA. We find that apparently the BHs gain mass much stronger than the host galaxies by a factor of $\sim$30. 
 These relative growth rates are broadly consistent with that seen in obscured AGN \citep{Netzer(2009),Silverman(2009)}. 
Such an offset implies that the typical duty cycle of an AGN (see \cite{Martini(2004)} for an overview) during which it can grow its BH mass efficiently must be short enough (typically $~10^7-10^8$~yr) to prevent a significant vertical movement in 
the black hole mass - bulge mass plane.  If the growth rates are extrapolated over a period of 1Gyr, the host galaxies do not gain much stellar mass from the present level of star formation.  
As previously mentioned, only one object (ID-333) shows a possible
major merger due to the presence of a more massive but inactive companion. While some objects show signs of minor merging activity (i.e. ID-712,ID-271), and we cannot exclude that we miss further minor merger events due to their low surface brightness, the stellar mass gain is 
expected to be low. Assuming the current growth rates and ignoring a possible major merger, all galaxies except one would need more than $>\sim1$Gyr to move more than 0.3 dex 
in the $M_\mathrm{BH}$-$M_\mathrm{*,Total}$. Therefore, we do not expect much evolution over the next 1Gyr for the majority of our sample.

\section{Summary}

We have performed a detailed analysis of a sample of 18 type 1 AGN host galaxies at $0.5<z<1.2$ to estimate their stellar mass content and explore the relation between the mass of the central BH and the mass of the host galaxy.  Our sample is of moderate-luminosity due to a selection based initially on their X-ray emission as detected with the Extended $Chandra$ Deep Field - South Survey.  This results in a sample having black hole masses below the knee of the black hole mass function thus mitigating biases \citep{Lauer(2007)} seen in other samples to date.  For the chosen redshift range, HST imaging is available with at least two filters that bracket the 4000~\AA\, break thus providing reliable stellar mass estimates of the host galaxy by accounting for both young and old stellar populations.  We have estimated bulge masses for all galaxies through either direct decomposition of the imaging data into a bulge or bulge plus disk component, or through simulations where artificial host galaxies with different B/T ratios are compared to single Sersic fits of the host galaxy. We are now able to look separately into their relation of the BH mass with either total stellar mass content or bulge mass after the contribution from the disk is removed. 

We find that the relation between $M_\mathrm{BH}$ and $M_\mathrm{*,Total}$ is in very good agreement with the local \ratio which has been reported by several studies so far.  
From our morphological analysis and decomposition of bulge and disk components, we can quantify the fraction of bulge dominated objects with $B/T>0.5$ to be 50\% while 72\% of the sample shows the presence of a disk component which is a significantly higher fraction than for a stellar mass matched local AGN sample.  Even though the bulge mass is shifted towards lower masses given their BH mass in some cases, we find that $\sim$80\% of the sample is in agreement with the local $M_\mathrm{BH}$ and $M_\mathrm{*,Bulge}$
relation given their $1\sigma$ error bars. We further compare the growth rates of the host galaxy and their BHs and find that assuming the present SFR and accretion rates (while ignoring possible 
major merger events), only one AGN in our sample would move more than 0.3 dex over the next 1Gyr. We highlight that bulge dominated galaxies are well in place at $z\sim1$ on the local \ratio relation.  There is a significant 
fraction (20\%) of our sample that is disk dominated and above the local relation which is not seen in either local inactive or active galaxy samples. For these galaxies to grow their bulges and align themselves on the local relation a physical mechanism is likely needed to redistribute their stars. While
mergers may play a role, it is not yet clear whether this is the dominant process. 

\acknowledgments

The authors fully appreciate the comments given by an anonymous referee that improved the paper and useful discussions with Tommaso Treu and Charles Steinhardt.  This work was supported by the World Premier International Research Center Initiative (WPI Initiative), MEXT, Japan.

\bibliography{reference}   

\end{document}